\documentclass{JHEP3}
\usepackage{amsmath}
\usepackage{amsfonts}
\usepackage{axodraw}
\usepackage[latin1]{inputenc}

\usepackage{cite}
\usepackage{epsfig}
\usepackage{xspace}
\usepackage{graphics}
\usepackage{inputenc}

\def\hide#1{}

\newcommand{\pythia}{P\protect\scalebox{0.8}{YTHIA}\xspace}

\newcommand{\ariadne}{A\protect\scalebox{0.8}{RIADNE}\xspace}

\newcommand{\done}[1]{}

\def\mrm#1{\mathrm{#1}}

\def\f2d3{\ensuremath{F_2^{\mrm{D}3}}}

\providecommand{\eqref}[1]{eq.~(\ref{#1})\xspace}
\renewcommand{\eqref}[1]{eq.~(\ref{#1})\xspace}

\newcounter{aenumct}

\newcounter{ienumct}

{\end{list}}

\newcounter{enumct}
\renewenvironment{enumerate}{\begin{list}{\arabic{enumct}.}%
{\usecounter{enumct}\setlength{\topsep}{1mm}%
\setlength{\partopsep}{1mm}\setlength{\itemsep}{0mm}%
\setlength{\parsep}{1mm}}}{\end{list}}

\usepackage{cite}

\newcommand{\dipsy}{{\small DIPSY}\xspace}

\newcommand{\be}[0]{\ensuremath{\beta}}

\newcommand{\ga}[0]{\ensuremath{\gamma}}

\newcommand{\de}[0]{\ensuremath{\delta}}
\newcommand{\al}[0]{\ensuremath{\alpha}}

\newcommand{\unit}[1]{\ensuremath{\left(1 - e^{#1} \right)}}

\newcommand{\All}[0]{\ensuremath{R}}
\newcommand{\all}[0]{\ensuremath{r}}

\newcommand{\Last}[0]{\ensuremath{L}}
\newcommand{\last}[0]{\ensuremath{l}}

\newcommand{\lastv}[0]{\ensuremath{{V_l}}}

\newcommand{\Cas}[0]{\ensuremath{P}}
\newcommand{\cas}[0]{\ensuremath{p}}
\newcommand{\Casv}[0]{\ensuremath{{V_P}}}
\newcommand{\casv}[0]{\ensuremath{{V_p}}}

\newcommand{\nn}[0]{\ensuremath{\ga}}
\newcommand{\yn}[0]{\ensuremath{\de}}

\author{Christoffer Flensburg, Gösta Gustafson, and Leif L\"onnblad\\
  Dept.~of Astronomy and Theoretical Physics, Lund University, \\
  Sölvegatan 14A, Lund, Sweden \vspace{2mm}\\
  E-mail: christoffer.flensburg@thep.lu.se, gosta.gustafson@thep.lu.se,
  leif.lonnblad@thep.lu.se} 

\abstract{In this paper we describe a formalism for generating exclusive final
states in diffractive excitation, based on the optical analogy where
diffraction is fully determined by the absorption into inelastic channels.
The formalism is based on the Good--Walker formalism for diffractive
excitation, and it is assumed that the virtual parton cascades represent
the diffractive eigenstates defined by a definite absorption amplitude.
We emphasize that, although diffractive excitation is basically a
quantum-mechanical phenomenon
with strong interference effects, it is possible to calculate the different
interfering components to the amplitude in an event generator, add them and thus
calculate the reaction cross section for exclusive diffractive final states.
The formalism is implemented in the DIPSY event generator, introducing no 
tunable parameters beyond
what has been determined previously in studies of non-diffractive events.
Some early results for DIS and proton-proton collisions are presented, and 
compared to experimental data.}

\title{Exclusive final states in diffractive excitation\footnote{Work
    supported in parts by the Swedish research council (contracts
    621-2009-4076 and 621-2010-3326).}}


\preprint{LU-TP 12-11\\
October 2012\\
\texttt{arXiv}: 1210.2407}

\begin{document}

\section{Introduction}

Although most events in high energy collisions have a continuous distribution
of hadrons, events with rapidity gaps are not rare; they contribute of the
order 10\% both in DIS at \textsc{Hera} \cite{Adloff:1997sc, Chekanov:2005vv}, 
and in $p\bar{p}$ 
collisions at the CERN $Sp\bar{p}S$ collider \cite{Bernard:1985kh,
Ansorge:1986xq} and the Tevatron \cite{Abe:1993wu}. 
These events have commonly been interpreted as the shadow of absorption into 
inelastic channels, in analogy with diffraction in optics.
In this view diffraction is expected to be a quantum-mechanical phenomenon,
to be analyzed at amplitude level, and not directly treatable by
semiclassical probabilistic methods.

In the Regge formalism the absorption is represented by cut pomeron or
reggeon diagrams. Regge diagrams, cut in between two exchanged
pomerons or reggeons, represent elastic scattering, while diffractive
excitation originates from triple pomeron diagrams, cut through one of
the pomerons \cite{Mueller:1970fa, Detar:1971gn}.  At high energies
multi-Regge diagrams, where the relative contributions from cut and
uncut diagrams are given by the AGK cutting rules
\cite{Abramovsky:1973fm}, are important.  Recent analyses within this
formalism include work by Kaidalov and coworkers
\cite{Kaidalov:2009hn}, Ostapchenko \cite{Ostapchenko:2010vb}, the
Durham \cite{Ryskin:2011qe, Ryskin:2012ry}, and the Tel Aviv
groups\cite{Gotsman:2009xm, Gotsman:2012rq, Gotsman:2012rm}.
These analyses give results for inclusive diffractive cross sections and
distributions in $d\sigma/d M_X^2$, but do not give information about
details in the diffractive final states. (Ref.~\cite{Kaidalov:2009hn} makes the 
extra assumption that a pomeron interacts like a $q\bar{q}$ pair.) A scheme for
generation of exclusive diffractive final states is outlined in
ref.~\cite{Martin:2012nm}, but as far as we know, no results are presented
yet. We note that, in contrast to our approach which is based on amplitudes, 
in this scheme diffractive events are generated as a semi-classical 
probabilistic process.

Diffractive excitation has also been described within the Good--Walker
formalism \cite{Good:1960ba}, where it is the result of differences in
absorption probability between different components of the projectile
wavefunction.  In the Regge approaches mentioned above, this formalism
was used only for low mass excitation. It was, however, early proposed
by Miettinnen and Pumplin \cite{Miettinen:1978jb}, that fluctuations
between parton cascades with different absorption probability also can
describe excitation to high masses.  In QCD a high energy proton is
visualized as a virtual cascade, with partons filling the rapidity
range between the proton rest frame and the observer. This implies
that the proton wavefunction also contains components with high masses
$\sim\! \exp(rapidity\,\,range)$. In high energy collisions the
cascades are determined by the BFKL dynamics, which has a stochastic
nature with large fluctuations, and in the dipole formulation these
cascades are also eigenstates of the interaction
\cite{Mueller:1989st}.  These fluctuations were studied by Hatta
\emph{et al.} \cite{Hatta:2006hs} in an analysis of diffractive
excitation in DIS at very high energies, within the saturated region
where $Q^2< Q_s^2$. This approach to diffractive excitation has also
been exploited by the Lund group within the dipole cascade model
implemented in the DIPSY Monte Carlo \cite{Avsar:2007xg,
  Flensburg:2010kq}, with applications to DIS and $pp$ collisions at
present day collider energies.
It is the aim of this paper to generalize this approach to calculate
amplitudes for transitions to exclusive final states.

In both the triple-Regge and the Good--Walker formalisms the diffractive 
amplitude is via the optical theorem determined by the inelastic cross section. 
At high energies inelastic events result from gluon exchange, which causes 
colour connections between projectile and target,
and the diffractive amplitude is represented by the (uncut)
perturbative BFKL pomeron formed by a two-gluon ladder.
The stochastic nature of the BFKL pomeron is also present in the AGK cutting
rules. In ref.~\cite{Gustafson:2012hg} it is argued that the triple-pomeron and 
Good--Walker formalisms only represent different views of the same
phenomenon. 
This idea is also supported by the fact that the bare pomeron in the 
DIPSY model also reproduces the
triple-Regge form for the diffractive excitation cross section
\cite{Flensburg:2010kq}.

In the ``colour reconnection'' approach it is assumed that the colour
exchange from an initial hard subcollision can be neutralized through the
mediation of subsequent soft gluons \cite{Edin:1995gi, Pasechnik:2010cm}. 
In this picture the cross section for
gap events is determined by a reconnection probability. This can be
tuned to fit experimental data, but is in this picture not dynamically fixed 
by the total inelastic cross section, and the relation to diffraction in
optics is not clear. 

Of particular interest in analyses of diffractive excitation are the
increased effects of saturation at higher energies.
In DIS diffractive excitation of the
photon has been calculated from elastic scattering of $q\bar{q}$ and $q\bar{q}g$
states. Golec-Biernat and W\"usthoff \cite{GolecBiernat:1999qd,GolecBiernat:2001mm} have emphasized that
 the coupling of a virtual photon is
significant to large $q\bar{q}$ dipoles with highly asymmetric energies, 
but that the coupling to the proton is suppressed for dipoles larger than
the saturation scale $R_0\sim 1/Q_s(x)$. This implies on one hand a suppression
of the diffractive cross section, and on the other hand that the interaction
is not dominated by soft interaction, but can be treated by perturbative QCD.
This argument should also be applicable for diffraction in $pp$
collisions, suggesting that also this can be treated by perturbative methods.

In $pp$ collisions the absorptive cross section is larger than in DIS.
In the Good--Walker formalism diffractive excitation is determined by
the fluctuations in absorption probability, which become small when the
black disk limit is approached. Thus
diffraction is dominated by elastic scattering, when the
saturation effects become large at higher energy, in particular for small
impact parameters. In the Regge formalism these saturation effects
 are represented by ``enhanced diagrams'', which interfere destructively and 
reduce the probability for a rapidity gap. Also in the colour reconnection
formalism, the large number of partons participating in the collision tends to
fill in a potential gap. In the analysis by Goulianos \cite{Goulianos:1995vn}, 
these saturation effects are described by a ``renormalized pomeron flux'', which
starts to become noticeable in $pp$ collisions around
$\sqrt{s} \approx 20$ GeV.

Most of the analyses cited above estimate inclusive distributions in 
$d\sigma/d M_X^2$ (or $d\sigma/d \eta_{\mathrm{max}}$) or semi-inclusive hard 
processes. To understand the properties of \emph{exclusive final states}
additional information or assumptions are needed. At lower masses, $M_X$, 
diffractively excited pions or protons
fragment in a string-like manner; an excited pion is similar to an
$e^+e^-$-annihilation event \cite{Adamus:1987kf}, while an excited proton has similarities to
DIS \cite{Smith:1985fa, Smith:1985vh}. These observations are in agreement
with the model by Donnachie
and Landshoff, in which the pomeron is assumed to interact 
with the quarks in the target like a photon
\cite{Donnachie:1984xq}. 

For higher masses, the limited acceptance of the detectors at 
the CERN $Sp\bar{p}S$ collider and the 
Tevatron implies that the experimental data 
usually do not cover the full phase space, and the experimental knowledge
about the properties of high mass diffractive final states is therefore also 
limited. At the CERN $Sp\bar{p}S$ collider rapidity distributions
were measured by the UA5 collaboration~\cite{Ansorge:1986xq}. The result was 
consistent with a $p_\perp$-limited
fragmentation, although there was here no experimental information about
transverse momenta. Similar results were obtained by the UA4 collaboration
\cite{Bernard:1985kh}, which pointed out the similarity between a diffractive 
system and a non-diffractive $pp$ collision at $\sqrt s =M_X$. 
At the \textsc{Hera} detectors the coverage of the 
photon fragmentation end 
is quite good, although an excited proton is mainly outside the acceptance.
Here the diffractively excited photon gives higher multiplicity and lower thrust
than an $e^+e^-$-annihilation event with the same mass \cite{Adloff:1997nn}.

High $p_\perp$ jets in diffractive states were observed 
by the UA8 collaboration \cite{Bonino:1988ae}, but the limited acceptance did
not allow a 
detailed study of the diffractive state, \emph{e.g.} of how the recoil was 
distributed. 
``Hard diffraction'' with high $p_\perp$ jets have later also been observed at
the Tevatron \cite{Affolder:2000vb, Abbott:1999km} and at 
\textsc{Hera}~\cite{:2007yw, Aaron:2011mp}. Due to the implications for
detection of exclusive central Higgs production, the observation of hard central
production with two rapidity gaps at the Tevatron, \emph{e.g.} of 
dijets \cite{Aaltonen:2007hs} and $e^+e^-$ pairs
\cite{Abulencia:2006nb}, has gained special interest.

 Hard diffraction has frequently been analyzed within 
the Ingelman-Schlein formalism
\cite{Ingelman:1984ns}, assuming that the pomeron interacts as composed of partons, in a way
similar to a hadron. The diffractive interaction is expressed
in terms of a flux factor $f_{\mathbb{P}p}(x_\mathbb{P})$, describing the
probability to find a pomeron with
energy fraction $x_\mathbb{P}$ in the proton, times the inelastic interaction
between the pomeron and the projectile proton or virtual photon. In this way 
analyses of \textsc{Hera} data for both hard
and soft diffraction have been described by a common set of ``pomeron structure
functions'' $f_{q/g,\mathbb{P}}(\beta)$, which have been fitted to a DGLAP evolution scheme 
\cite{:2009qja}. These analyses indicate a dominant gluonic component in the pomeron.
This leads to more complicated final states than the straight strings seen at
lower masses, in agreement with the observations in refs.~\cite{Bernard:1985kh}
and \cite{Adloff:1997nn} mentioned above. 

This approach is also implemented in several Monte Carlo (MC) event
generators used in analyses of experimental data, \emph{e.g.}
\textsc{Pompyt} \cite{Pompyt:1996}, \textsc{Rapgap}
\cite{Jung:1993gf}, and \textsc{Pythia8} \cite{Sjostrand:2007gs}.  In
these models the result depends on pomeron flux factors and pomeron
structure functions, which have to be fitted to data. Here the effects
of saturation have to be included in the flux factor, or in ``gap
survival factors''.

The aim of this paper 
is to present a formalism, in which the cross section for exclusive
diffractive final states can be calculated directly based on the dynamical 
features of small
$x$ evolution and saturation. The formalism is an extension of the Lund
Dipole Cascade model implemented in the DIPSY MC
\cite{Avsar:2005iz,Flensburg:2011kk}, 
which is based on BFKL evolution, including essential non-leading effects,
and saturation within the cascade evolution. Hence the model is based on
perturbative QCD, which might be motivated by Golec-Biernat--W\"usthoff's
argument that gluons below the saturation scale $Q_s$ are suppressed at small
$x$. Diffractive excitation is calculated within the Good--Walker formalism 
with no additional free parameters. Results for inclusive diffractive cross
sections have been presented previously \cite{Avsar:2007xg, Flensburg:2010kq},
the generation of exclusive non-diffractive final states was discussed in
ref.~\cite{Flensburg:2011kk}, 
and the model is here extended to describe exclusive final states in
diffractive excitation.

We want to emphasize that, although in the optical analogy diffractive 
excitation is basically a quantum-mechanical phenomenon with strong 
interference effects, it is in our formalism possible to calculate the different
interfering contributions to the amplitude in the DIPSY MC, add them with
their proper signs, and thus calculate the reaction cross section. 
This distinguishes our formulation from other approaches, in which
final states are generated as a semi-classical probabilistic process.

The outline of the paper is as follows: In section 2 we review shortly the
eikonal approximation and the Good--Walker formalism. The dipole cascade model
is described in section 3, and in section 4 we discuss how this model can be
extended to describe exclusive final states in diffractive excitation. In
section 5 we discuss the implementation in the DIPSY event generator, and in
section 6 we present some early results. Some possible future developments
are discussed in section 7, followed by our conclusions in section 8.

\section{Diffractive excitation in the Good--Walker picture}

\subsection{The Good--Walker formalism}
\label{sec:subGW}

In the Good--Walker formalism \cite{Good:1960ba}, diffraction in
hadronic collisions is analogous to diffraction in optics, and a
result of absorption and unitarity. Unitarity constraints and
saturation are most easily treated by the eikonal formalism in impact
parameter space.

Assume that the elastic scattering is driven by absorption into a
number of inelastic states $n$, with Born amplitudes $\sqrt{2f_n}$. In
the eikonal approximation \cite{Glauber:1957} the absorption
probability, $P_{\mathrm{abs}}$, is given by the inelastic cross
section
\begin{equation}
P_{\mathrm{abs}}=d\sigma_{\text{inel}}/d^2b =  \, 1-e^{-2F}, \,\,\mathrm{where} \,\,F\equiv \sum f_n.
\label{eq:sigmaabs}
\end{equation}
Thus the probability for the projectile not to be absorbed into any of the
inelastic states is given by $e^{-2F}$.

For a structureless projectile unitarity then gives the $S$-matrix
\begin{equation}
S = e^{-F}=e^{-\sum f_n}.
\label{eq:S}
\end{equation} 
Thus the amplitude for elastic scattering is given by
\begin{equation}
T = 1-S=1-e^{-F} \label{eq:unit},
\end{equation} 
which satisfies the optical theorem $T=\frac{1}{2}(T^2+P_{\mathrm{abs}})$.
Note that we have here defined the amplitude $T$ without the conventional
factor $i$, so that $T$ becomes real. The elastic cross section is given by
$d\sigma_{\text{el}}/d^2b \,\, = \,  T^2$, 
and adding the inelastic cross section, the total cross section is given by 
$d\sigma_{\text{tot}} / d^2b=2(1-e^{-F})= 2T$.

If the projectile has an \emph{internal structure}, it can be excited in a
diffractive scattering event. This implies that the mass eigenstates $\Psi_k$
(asymptotic incoming and outgoing states) can
differ from the diffractive eigenstates, \emph{i.e.} eigenstates to the 
$S$-matrix.
We denote the diffractive eigenstates $\Phi_i$, with eigenvalues given by 
$S \Phi_i=e^{-F_i} \Phi_i$. They can be
absorbed with probability $1-e^{-2F_i}$, and have elastic
scattering amplitudes $\langle \Phi_i|T|\Phi_j\rangle=\delta_{ij} T_i$, with $T_i=1-e^{-F_i}$. 

The diffractive eigenstates 
can be written as linear combinations of the mass
eigenstates, 
\begin{equation}
\Phi_i= \sum_k  c_{ik} \Psi_k,
\label{eq:cki}
\end{equation}
where $c_{ik}$ is a unitary matrix (thus $\Psi_k=\sum c^\dagger_{ki}\Phi_i$), and the incoming 
state is given by $\Psi_{\text{in}}=\Psi_0$. 
The elastic amplitude is then given by the average over the diffractive eigenstates:
\begin{equation} 
\langle \Psi_0 | T | \Psi_0 \rangle = \sum |c_{i0}|^2 T_i 
= \langle T \rangle,
\label{eq:Tel}
\end{equation}
with the elastic cross section
\begin{equation}
d \sigma_{\text{el}}/d^2 b = \left( \sum |c_{i0}|^2 T_i \right)^2 = \langle T
\rangle ^2.
\label{eq:eikonalel}
\end{equation}
The amplitude for diffractive transition to the mass eigenstate $\Psi_k$
becomes
\begin{equation}
\langle \Psi_{k} | T | \Psi_0 \rangle = \sum_i  c^\dagger_{ki} T_i c_{i0},
\label{eq:Tik}
\end{equation}
which gives a total diffractive cross section (including elastic scattering)
\begin{equation}
d\sigma_{\text{diff}}/d^2 b
=\sum_k \langle \Psi_0 | T | \Psi_{k} \rangle \langle \Psi_{k} | T |
\Psi_0 \rangle =\langle T^2 \rangle.
\label{eq:GWsigmadiff}
\end{equation}
Subtracting the elastic scattering we find the cross section for 
diffractive excitation
\begin{equation}
d\sigma_{\text{diff ex}}/d^2 b  = d\sigma_{\text{diff}}/d^2 b - d \sigma_{\text{el}}/d^2 b =
\langle T^2 \rangle - \langle T \rangle ^2\equiv V_T,
\label{eq:eikonaldiff}
\end{equation}
which thus is determined by the fluctuations in the scattering
process.

\subsection{Diffractive eigenstates}
\label{sec:eigenstates}

\begin{figure}
  \begin{center}
    \begin{minipage}[]{0.23\linewidth}
      \begin{center}
        \includegraphics[width=0.75\linewidth]{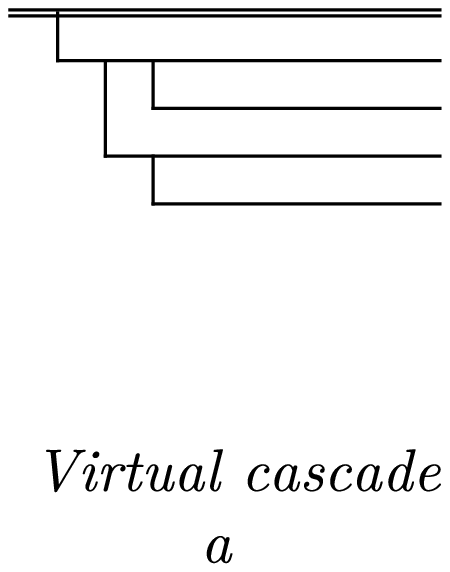}
      \end{center}
    \end{minipage}
    \begin{minipage}[]{0.23\linewidth}
      \begin{center}
        \includegraphics[width=0.75\linewidth]{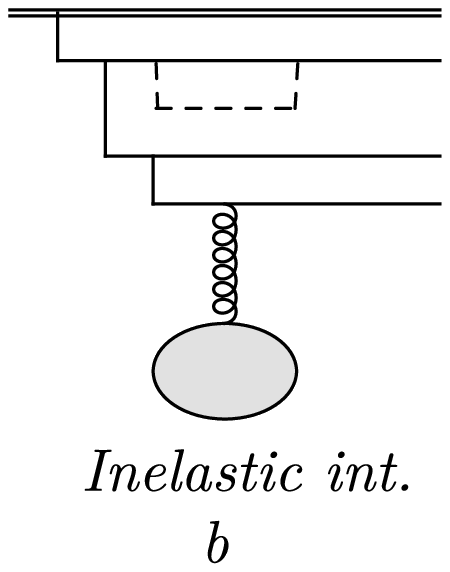}
      \end{center}
    \end{minipage}
    \begin{minipage}[]{0.23\linewidth}
      \begin{center}
        \includegraphics[width=0.75\linewidth]{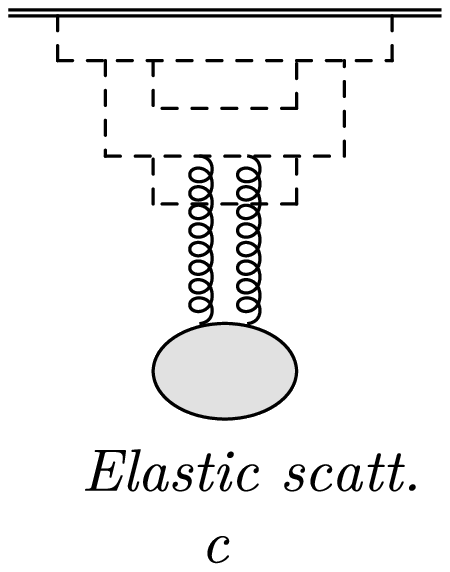}
      \end{center}
    \end{minipage}
    \begin{minipage}[]{0.23\linewidth}
      \begin{center}
        \includegraphics[width=0.75\linewidth]{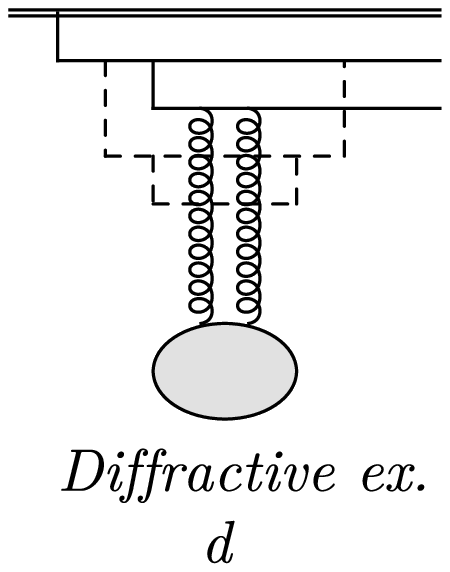}
      \end{center}
    \end{minipage}
  \end{center}
\caption{\label{fig:eigenstates} (a) An example of a parton (or dipole) 
  cascade evolved in
  rapidity. (b) The exchange of a gluon gives rise to an inelastic
  interaction. (c) Elastic scattering is obtained from coherent scattering of
  different partons in different cascades, via the exchange of two gluons. (d) 
  Diffractive excitation is obtained when the result of the two-gluon exchange
  does not correspond to the coherent initial proton state. Dashed
  lines indicate virtual emissions, which are not present in the
  final state.}
\end{figure}

The basic assumption in our earlier analyses of diffractive excitation in 
refs.~\cite{Avsar:2007xg, Flensburg:2010kq}, is
that the diffractive eigenstates correspond to
parton cascades, which can come on shell through interaction with the target.
As mentioned in the introduction, this was also the assumption in the early
work by Miettinen and Pumplin \cite{Miettinen:1978jb}, and a similar approach
has been used by Hatta \textit{et al.} \cite{Hatta:2006hs}.
This means that the factors $c_{i0}$ in eq.~(\ref{eq:cki}) are given by the evolution
of the virtual cascade. The process is illustrated in
fig.~\ref{fig:eigenstates}. Fig.~\ref{fig:eigenstates}$a$ shows the
virtual cascade before the collision, and fig.~\ref{fig:eigenstates}$b$ 
illustrates an inelastic interaction, where gluon exchange gives a colour
connection between 
the projectile and the target. This implies that the beam is absorbed with probability
$1-e^{-2F_i}$. In this and the following figures, solid lines represent real
emissions present in the final state, while a dashed line corresponds to a
virtual emission, which did not come on shell via the interaction with the
target, and therefore is reabsorbed in the cascade. 
Fig.~\ref{fig:eigenstates}$c$ shows an 
elastic interaction. Via the optical theorem the elastic amplitude is given by the inelastic 
cross section, represented by the square of the diagram in fig.~\ref{fig:eigenstates}b,
which corresponds to the exchange of two gluons. Note that
the elastic amplitude in eq.~(\ref{eq:Tel})
is the result of \emph{coherent} interaction of all partons in all
possible cascades. Fig.~\ref{fig:eigenstates}$d$, finally, shows the 
contribution to the scattered beam, which is orthogonal to the incoming
state. This corresponds to
diffractive excitation, with the amplitude given by eq.~(\ref{eq:Tik}).
The lines can 
symbolise gluons in a traditional cascade, or dipoles in a dipole cascade.
When the diagram in fig.~\ref{fig:eigenstates}$d$ is squared, it is consistent
with Mueller's triple Regge formalism \cite{Mueller:1970fa}.

\section{The dipole cascade model}
\label{sec:dipcasc}
\subsection{Mueller's dipole model}

\begin{figure}
\begin{center}
\scalebox{0.8}{\mbox{
\begin{picture}(340,80)(0,0)
\Vertex(10,80){2}
\Vertex(10,0){2}
\Text(5,80)[]{$\mathrm{x}$}
\Text(5,0)[]{$\mathrm{y}$}
\Line(10,80)(10,0)
\LongArrow(30,40)(60,40)
\Vertex(100,80){2}
\Vertex(100,0){2}
\Vertex(120,50){2}
\Text(95,80)[]{$\mathrm{x}$}
\Text(95,0)[]{$\mathrm{y}$}
\Text(128,50)[]{$\mathrm{z}$}
\Line(100,80)(120,50)
\Line(120,50)(100,0)
\LongArrow(140,40)(170,40)
\Vertex(205,80){2}
\Vertex(225,50){2}
\Vertex(233,30){2}
\Vertex(205,0){2}
\Text(200,80)[]{$\mathrm{x}$}
\Text(200,0)[]{$\mathrm{y}$}
\Text(233,50)[]{$\mathrm{z}$}
\Text(241,30)[]{$\mathrm{w}$}
\Line(205,80)(225,50)
\Line(225,50)(233,30)
\Line(233,30)(205,0)
\LongArrow(255,40)(285,40)
\Line(310,80)(320,70)
\Line(320,70)(302,63)
\Line(302,63)(325,58)
\Line(325,58)(330,50)
\Line(330,50)(320,43)
\Line(320,43)(338,30)
\Line(338,30)(310,35)
\Line(310,35)(310,0)
\Vertex(310,80){2}
\Vertex(320,70){2}
\Vertex(302,63){2}
\Vertex(325,58){2}
\Vertex(330,50){2}
\Vertex(320,43){2}
\Vertex(338,30){2}
\Vertex(310,35){2}
\Vertex(310,0){2}
\end{picture}
}}
\end{center}
\caption{Gluon emission splits the dipole into two dipoles. Repeated
emissions give a cascade, which produces a chain of dipoles.}
\label{fig:dipolesplit}
\end{figure}
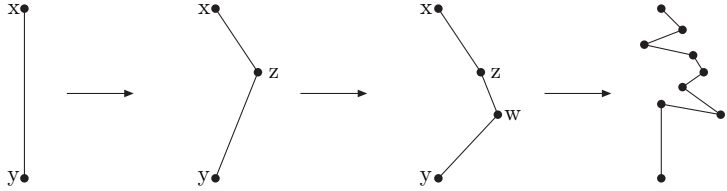

Mueller's dipole cascade model
\cite{Mueller:1993rr,Mueller:1994jq,Mueller:1994gb} is a formulation
of LL BFKL evolution in transverse coordinate space. 
Gluon radiation from the colour charge in a parent quark or gluon is screened 
by the accompanying anticharge 
in the colour dipole. This suppresses emissions at large transverse separation,
which corresponds to the suppression of small $k_\perp$ in BFKL.
For a dipole with charges in transverse points $\mathbf{x}$ and $\mathbf{y}$,
the probability per unit rapidity ($Y$) 
for emission of a gluon at transverse position $\mathbf{z}$ is given by
\begin{eqnarray}
\frac{d\mathcal{P}}{dY}=\frac{\bar{\alpha}}{2\pi}d^2\mathbf{z}
\frac{(\mathbf{x}-\mathbf{y})^2}{(\mathbf{x}-\mathbf{z})^2 (\mathbf{z}-\mathbf{y})^2},
\,\,\,\,\,\,\, \mathrm{with}\,\,\, \bar{\alpha} = \frac{N_c\alpha_s}{\pi}.
\label{eq:dipkernel1}
\end{eqnarray}
As shown in fig.~\ref{fig:dipolesplit}
the dipole is split into two dipoles, which
(in the large $N_c$ limit) emit new gluons independently. The result is a
cascade, giving a dipole chain where the number of dipoles grows exponentially
with $Y$.

\begin{figure}
\begin{center}
\scalebox{1.3}{
\begin{picture}(180,80)(0,0)

\Text(29,28)[tl]{{\footnotesize $y_i$}}
\Text(29,42)[bl]{{\footnotesize $x_i$}}
\Text(52,44)[br]{{\footnotesize $y_j$}}
\Text(52,29)[tr]{{\footnotesize $x_j$}}
\Vertex(150,32){1}
\Vertex(130,30){1}
\Vertex(130,40){1}
\Vertex(151,43){1}
\Vertex(50,32){1}
\Vertex(30,30){1}
\Vertex(30,40){1}
\Vertex(51,43){1}

\Vertex(10,20){1}
\Vertex(18,70){1}
\Vertex(70,20){1}
\Vertex(60,65){1}
\Vertex(20,50){1}
\Vertex(28,57){1}
\Vertex(60,10){1}
\Vertex(60,52){1}

\Vertex(110,20){1}
\Vertex(118,70){1}
\Vertex(170,20){1}
\Vertex(160,65){1}
\Vertex(120,50){1}
\Vertex(128,57){1}
\Vertex(160,10){1}
\Vertex(160,52){1}
\LongArrowArcn(90,20)(20,120,60)

\Line(10,20)(30,30)
\ArrowLine(30,40)(30,30)
\Line(30,40)(20,50)
\Line(20,50)(28,57)
\Line(28,57)(18,70)

\Line(70,20)(60,10)
\Line(60,10)(50,32)
\ArrowLine(50,32)(51,43)
\Line(51,43)(60,52)
\Line(60,52)(60,65)

\ArrowLine(130,30)(110,20)
\ArrowLine(120,50)(130,40)
\Line(120,50)(128,57)
\Line(128,57)(118,70)

\Line(170,20)(160,10)
\ArrowLine(160,10)(150,32)
\ArrowLine(151,43)(160,52)
\Line(160,52)(160,65)

\ArrowLine(150,32)(130,30)
\ArrowLine(130,40)(151,43)

\end{picture}
}
\end{center}
\caption{An interaction between a dipole in the projectile and another in the
  target due to gluon exchange gives a recoupling of the dipole chains.}
\label{fig:dipscatt}
\end{figure}
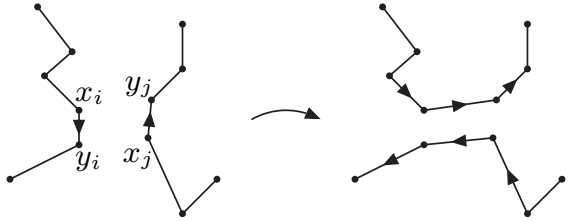

When two cascades collide, a pair of dipoles with coordinates 
$(\mathbf{x}_i,\mathbf{y}_i)$  and $(\mathbf{x}_j,\mathbf{y}_j)$, in 
the projectile and target respectively, can interact 
via gluon exchange with the probability $2f_{ij}$, where
\begin{equation}
  f_{ij} = f(\mathbf{x}_i,\mathbf{y}_i|\mathbf{x}_j,\mathbf{y}_j) =
  \frac{\alpha_s^2}{8}\biggl[\log\biggl(\frac{(\mathbf{x}_i-\mathbf{y}_j)^2
    (\mathbf{y}_i-\mathbf{x}_j)^2}
  {(\mathbf{x}_i-\mathbf{x}_j)^2(\mathbf{y}_i-\mathbf{y}_j)^2}\biggr)\biggr]^2.
\label{eq:dipamp}
\end{equation}
We note here that the interaction 
probability goes to zero for a small dipole. This implies that the singularity
in the production probability for small dipoles in 
eq.~(\ref{eq:dipkernel1}) does not give infinite cross sections.
We note also that gluon exchange means exchange of colour between the two 
cascades. This implies a reconnection of the dipole chains, as shown in
fig.~\ref{fig:dipscatt}, and the formation of dipole chains connecting the
projectile and target remnants. 

\begin{figure}
\begin{center}
\includegraphics[width=.65\linewidth]{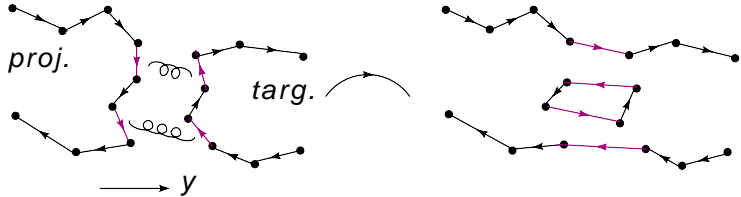}
\end{center}
\caption{Double interaction results in a dipole loop, corresponding to a pomeron
  loop.}
\label{fig:loop}
\end{figure}

In Mueller's model the constraints from unitarity are satisfied using the 
eikonal formalism.
When more than one pair of dipoles interact, colour loops are formed, as shown
in fig.~\ref{fig:loop}. 

\subsection{The Lund dipole cascade model DIPSY}
\label{sec:lundcascade}
It is well known that non-leading effects are very important in BFKL
evolution. Another problem is that Mueller's model does not include
pomeron loops within the cascade evolution.  The Lund model
\cite{Avsar:2005iz, Avsar:2006jy, Flensburg:2008ag, Flensburg:2011kk}
is a generalization of Mueller's model, which also includes NLL BFKL
effects, nonlinear (colour-suppressed) effects in the evolution, and
confinement effects.

\textbf{a) Beyond LL BFKL}

The NLL corrections to BFKL evolution have three major sources
\cite{Salam:1999cn}:

\emph {Non-singular terms in the splitting function:}
These terms suppress large $z$-values in the individual parton branchings. 
Most of this effect is taken care of by including energy-momentum
conservation. This is effectively taken into account by associating a
dipole with transverse size $r$ with a transverse momentum $k_\perp = 1/r$,
and demanding conservation of the light-cone momentum $p_+$ in every step
in the evolution. This gives an effective cutoff for small dipoles.

\emph {Projectile-target symmetry:}
A parton chain should look the same if generated from the target end
as from the projectile end. The corresponding corrections are also
called energy scale terms, and are essentially equivalent to the so
called consistency constraint \cite{Kwiecinski:1996td}. This effect is
taken into account by conservation of the negative
light-cone momentum components, $p_-$.

\emph {The running coupling:}
Following ref.~\cite{Balitsky:2008zzb}, the scale in the running coupling is 
taken as the largest transverse momentum in the vertex.

\textbf{b) Nonlinear effects in the evolution}

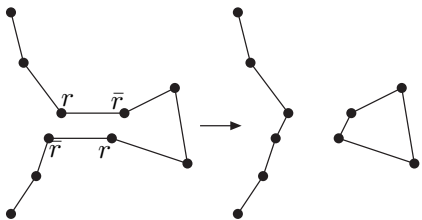
\begin{figure}
\begin{center}
\scalebox{0.95}{\mbox{
\begin{picture}(180,100)(0,0)
\Line(10,10)(20,25)
\Line(25,40)(20,25)
\Line(50,40)(80,30)
\Line(75,60)(80,30)
\Line(75,60)(55,50)
\Line(30,50)(15,70)
\Line(15,70)(10,90)
\LongArrow(85,45)(100,45)
\Line(100,10)(110,25)
\Line(115,40)(110,25)

\Line(140,40)(170,30)
\Line(165,60)(170,30)
\Line(165,60)(145,50)

\Line(120,50)(105,70)
\Line(105,70)(100,90)
\Vertex(10,10){2}
\Vertex(20,25){2}
\Vertex(25,40){2} 
\Vertex(50,40){2} 
\Vertex(80,30){2} 
\Vertex(75,60){2}
\Vertex(55,50){2} \Vertex(30,50){2} \Vertex(15,70){2} \Vertex(10,90){2}
\Vertex(100,10){2} \Vertex(110,25){2} \Vertex(115,40){2} \Vertex(120,50){2}
\Vertex(140,40){2} \Vertex(170,30){2} \Vertex(165,60){2} \Vertex(145,50){2}
\Vertex(105,70){2} \Vertex(100,90){2}
\Line(25,40)(50,40)
\Line(30,50)(55,50)
\Line(115,40)(120,50)
\Line(145,50)(140,40)
\Text(25,39)[tl]{$\bar{r}$}
\Text(50,37)[tr]{$r$}
\Text(30,53)[bl]{$r$}
\Text(55,52)[br]{$\bar{r}$} 
\end{picture}
}}
\end{center}
\caption{Two dipoles with the same colour form a colour octet, which may be
  better approximated by dipoles formed by the closet colour-anticolour
  pairs. This implies a recoupling of the dipole chains, which favours the
  formation of small dipoles. Thus the number of dipoles resolved by a
  probe with given resolution $Q^2$ is reduced.}
\label{fig:swing}
\end{figure}

As mentioned above, multiple interactions produce loops of dipole chains
corresponding to pomeron loops. Mueller's model includes all loops cut in 
the particular Lorentz frame used for the analysis, but not loops contained
within the evolution of the individual projectile and target cascades.  
As for dipole scattering
the probability for such loops is given by $\alpha_s$, and therefore
formally colour suppressed compared to dipole splitting, which is
proportional to $\bar{\alpha}=N_c \alpha_s/\pi$. These loops are therefore
related to the probability that two dipoles have the same colour. Two dipoles
with the same colour form a quadrupole field. Such a field may be better
approximated by two dipoles formed by the closest colour--anticolour
charges. This corresponds to a recoupling of the colour dipole chains. The
process is illustrated in Fig.~\ref{fig:swing}, and we call
it a dipole ``swing''.  With a weight for the swing which favours small
dipoles, we obtain an almost frame independent result.
The number of dipoles in the cascade is not reduced, and the saturation 
effect is a consequence of the smaller interaction probability for 
the smaller dipoles. Thus the number of dipoles (or gluons) resolved by a
probe with a given resolution $Q^2$, is reduced. In this way the swing also
generates effectively $2\rightarrow 1$, or in some cases a $2\rightarrow 0$,
transitions. 

\vspace{2mm}

\textbf{c) Confinement} 

As mentioned earlier, saturation effects imply that the cascade evolution 
is dominated by relatively small dipoles. However, although suppressed,
the rare large dipoles generated in a purely perturbative evolution with 
massless gluons give non-negligible effects, and eventually 
Froissart's bound will be violated \cite{Avsar:2008dn}.
Therefore confinement is also important, and is taken into account by 
giving the gluon an effective mass.

\subsection{Inclusive cross sections and non-diffractive final states}
\label{sec:cross-sections}

The Lund cascade model is implemented in a Monte Carlo event generator
called DIPSY, with applications to collisions between electrons,
protons, and nuclei. An incoming virtual photon is here treated as a
$q\bar{q}$ pair, with an initial state wavefunction determined by QED.
For an incoming proton we make an ansatz in form of an equilateral
triangle of dipoles, but after evolution the result is rather
insensitive to the exact form of the initial state.

The model reproduces well the inclusive total and elastic
cross sections in $pp$ collisions and DIS, as described in ref.~\cite{Flensburg:2008ag}. 
Diffractive cross sections are reproduced within the Good--Walker formalism
\cite{Flensburg:2010kq}. It should, however, here be noted that when the nonlinear effects from 
saturation are switched off, the result agrees with the triple-pomeron formula
for a bare pomeron with intercept $\alpha(0)=1.21$ and an approximately 
constant triple-pomeron coupling $g_{3P}=0.3\, \mathrm{GeV}^{-1}$. This result
is a consequence of the common underlying assumption, with diffraction
as the shadow of absorption.

The model is also generalized to describe the properties
of exclusive non-diffractive final states \cite{Flensburg:2011kk}. 
We here note that BFKL evolution properly reproduces inclusive observables.
For exclusive final states it is necessary to take into account colour 
coherence and angular ordering, as well as soft radiation, related to the 
$z=1$ singularity in the gluon splitting function. These effects are taken 
into account in the CCFM formalism \cite{Catani:1989sg, Ciafaloni:1987ur}, 
which also reproduces the BFKL result for inclusive cross sections.

An essential point is here the fact that the softer emissions in the CCFM 
formalism can be resummed, and the total cross section, as well as the structure 
of the final states, are fully determined by the "$k_\perp$-changing" gluons.
We denote the real emitted gluons in a ladder $q_i$ and the virtual links $k_i$, 
and momentum conservation then implies $\mathbf{k}_{\perp i-1}=
\mathbf{k}_{\perp i}+\mathbf{q}_{\perp i}$. For $k_\perp$-changing emissions
$k_{\perp i}$ is either much larger or much smaller than $k_{\perp i-1}$. This also 
means that $q_{\perp i} \approx \max(k_{\perp i}, k_{\perp i-1})$. These emissions
are called "primary gluons" in ref.~\cite{Andersson:1995ju}, and "backbone gluons" in 
ref.~\cite{Salam:1999ft}, and the weight for such a backbone chain is given by
\begin{equation}
\mathrm{weight} = \prod \bar{\alpha} \frac{d q_{\perp i}^2}{q_{\perp i}^2} d y_i.
\label{eq:backbone}
\end{equation}
As discussed in ref.~\cite{Gustafson:1999kh}, this feature also gives a
dynamical cutoff for 
small $q_\perp$, which grows slowly with energy, and gives a dynamical
description of the cutoff for hard subcollisions needed in event
generators like \textsc{Pythia}~\cite{Sjostrand:2007gs} or 
\text{Herwig}~\cite{Bahr:2008pv}. 

To generate the inelastic final states, we thus have to go through the 
following steps:
\begin{itemize}
\item Generate two dipole cascades for the projectile and the target.
\item Determine which dipoles in the projectile and target become colour connected
  via gluon exchange.
\item Extract the backbone chains from the cascades, and remove virtual branches,
which cannot come on shell by the interaction.
\item Add softer emissions as final state radiation, with appropriate Sudakov
form factors.
\item Hadronize the gluonic chains.
\end{itemize}
The results from MC simulations, presented in ref.~\cite{Flensburg:2011kk},
are able to give a fair  
reproduction of data for minimum bias and underlying events from LHC and the 
Tevatron, remembering that this implementation does not include
matrix element corrections for hard scattering or contributions from quarks.

\section{Exclusive final states in diffractive excitation}
\label{sec:finalstates}

\subsection{Basic formalism}
We will now discuss how to describe exclusive states in
diffractive excitation, within the dipole cascade model and the Good--Walker
formalism. As discussed in sec.~\ref{sec:eigenstates}, the partonic cascades are 
interpreted as the diffractive eigenstates. Thus the unitary matrix $c_{ik}$ in 
eq.~(\ref{eq:cki}) is represented by a unitary operator, describing the cascade 
evolution. The cascade is absorbed with probability $1-e^{-2F}$, and via
the optical theorem this gives an elastic scattering amplitude $T=1-e^{-F}$.
The hermitian conjugate matrix, $c^\dagger_{ki}$, in the amplitude  for diffractive 
transition to another mass eigenstate in eq.~(\ref{eq:Tik}), represents
an evolution backwards in rapidity, which can reabsorb
some of the emitted partons. As cascades with more partons generally are
absorbed with higher probability, this backward evolution does not always give the 
original proton back, but a different partonic system, interpreted as a possible mass 
eigenstate.  

The evolution in rapidity may also be interpreted as an evolution in time, from an
incoming state at $t=-\infty$ to the cascade present just before the collision at time $t=0$. 
Thus the states at $t=-\infty$ correspond to the mass eigenstates $\Psi_k$ in 
sec.~\ref{sec:subGW} (we include the possibility that this state does not
correspond to the ground state $\Psi_0$), while the evolved states at time $t=-\delta$ correspond to the
diffractive eigenstates $\Phi_i$. As the possible cascades are not a discrete set, we
describe the cascade evolution by a unitary operator $U(-\delta, -\infty)$, replacing the 
unitary matrix $c_{ik}$ in eq.~(\ref{eq:cki}). The interaction of the cascades with
the target is described by a diagonal operator $T$, which corresponds to an evolution 
operator $U(+\delta,-\delta)\equiv U_{\mathrm{int}}=\mathbb{I}-T$ (where
$\mathbb{I}$ denotes the unit matrix). The
transformation back to mass eigenstates, obtained by the inverse of the cascade evolution
operator, can then be interpreted as describing the evolution from time $t=+\delta$
to $t=\infty$, $U(+\infty,+\delta)=U(-\delta, -\infty)^{-1}=U(-\delta,
-\infty)^\dagger$.   
Thus the total $S$-matrix is given by
\begin{equation}
S=U(+\infty,+\delta)U(+\delta,-\delta)U(-\delta, -\infty)=U(-\delta, -\infty)^{-1}
U_{\mathrm{int}}U(-\delta, -\infty),
\end{equation}
with the diffractive final state given by
\begin{equation}
\Psi_{\mathrm{out}}=S \Psi_{\mathrm{in}}.
\end{equation} 
In order to simplify the subsequent equations, we here expressed the
relations in terms of the $S$-matrix, instead of the transition
amplitude $T$.  Note that as the diffractive scattering is the shadow
of absorption, this operator in not unitary.  In section
\ref{sec:combined} an extended formalism is discussed, in which the
non-diffractive states are included, and the $S$-matrix is a unitary
operator.

In order to illustrate the essential features in diffractive
excitation, we first study some simple toy models. The results will
help understanding the more realistic gluon cascades discussed in the
later subsections.  In the toy models in sec.~\ref{sec:toy}, and in
the continuous cascade in sec.~\ref{sec:cont}, the gluons are assumed
to interact individually, with no effects of screening from
neighbouring gluons. Such screening effects are taken into account in
the dipole cascade discussed in sec.~\ref{sec:dipole}. We will here
also assume that the cascade only includes gluon emission, and that
effects of saturation from gluons joining (sometimes referred to as
gluon recombination) can be neglected. Such effects are discussed in
secs.~\ref{sec:MCsat} and \ref{sec:swing}.

\subsection{Toy models}
\label{sec:toy}

\subsubsection{A two-particle system}
\label{sec:toy1}

We first study the simple case of a particle, the valence particle, which can 
emit
a single gluon at a fixed point in phase space. (This example is also directly
analogous to the 2-channel Good--Walker treatment of low mass excitation in
the work by the Tel Aviv or Durham groups.) The state with no emission and 
only the original
valence particle is denoted $|1,0\rangle$, and the state including the emitted gluon is
denoted $|1,1\rangle$. Thus the numbers indicate whether the two particles are 
present (1) or absent (0). At asymptotic times $t=\pm\infty$ these states are also mass eigenstates,
$|1,0\rangle_{\pm\infty}$ and $|1,1\rangle_{\pm\infty}$ respectively. 
During the evolution
up to the collision, the particle can emit a gluon. This is described by the 
unitary evolution operator $U(-\delta, -\infty)$, which contains a 
$|1,0\rangle \rightarrow |1,1\rangle$ transition. In the ($|1,0\rangle, |1,1\rangle$) space it can be
described by a unitary matrix (we choose phases such that $U$ is real):
\begin{equation}
U(-\delta,-\infty) = \left( \begin{array}{cc}
\al & -\be \\
\be & \al \end{array} \right), \qquad \al^2 + \be^2 = 1 . \label{eq:toy1U}
\end{equation}
Thus $\beta^2$ gives the probability for gluon emission, and $\alpha^2$ the
probability for no emission.

In the collision, which takes place in a short time interval
$(-\delta,+\delta)$, we assume that the valence particle and the gluon can be 
absorbed with probabilities $1-e^{-2f_0}$ and $1-e^{-2f_1}$ respectively. 
In this interaction the beam particle becomes colour-conected to the target,
and thus absorbed from the incoming beam. This
implies that the states $|1,0\rangle$ and $|1,1\rangle$ have probabilities $e^{-2f_0}$ and
$e^{-2(f_0+f_1)}$ \emph{not} to be absorbed. In accordance with the
optical theorem, the evolution operator 
describing the evolution of the diffractive eigenstates during the
collision, is given by (\emph{cf} eq.~(\ref{eq:S})):
\begin{equation}
U_{\mathrm{int}}=U(+\delta,-\delta)=\left(\begin{array}{cc}
e^{-f_0} & 0 \\
0 & e^{-f_0-f_1} \end{array} \right).
\label{eq:toy1Uint}
\end{equation}

Finally the system evolves from time $+\delta$ to $+\infty$, described
 by the hermitian conjugate of the unitary matrix
$U(-\delta, -\infty)$: 
\begin{equation}
U(+\infty,+\delta) = U^\dagger(-\delta,-\infty) =\left( \begin{array}{cc}
\al & \be \\
-\be & \al \end{array} \right)
\end{equation}
The $S$-matrix for the diffractive states is thus given by
\begin{eqnarray}
S=U( +\infty,+\delta) U_{\mathrm{int}} U(-\delta,-\infty)=\nonumber\\
\left( \begin{array}{cc}
e^{-f_0}(\alpha^2 +\beta^2\, e^{-f_1}), & -\alpha\beta\, e^{-f_0}(1-e^{-f_1}) \\
-\alpha\beta\, e^{-f_0}(1-e^{-f_1}), & e^{-f_0}(\beta^2 +\alpha^2\,
e^{-f_1}) \end{array} \right).
\label{eq:Stoy1}
\end{eqnarray}
The scattering matrix is given by $T= \mathbb{I}-S$, and from the above 
expression we can read off the scattering
amplitudes for an incoming single valence particle. Thus the amplitudes for
elastic scattering and diffractive excitation can be written as 
(recalling that $\al^2 + \be^2 = 1$)
\begin{eqnarray}
\mathrm{Elastic\,\, scattering}:&&T(|1,0\rangle \rightarrow |1,0\rangle) =1 -S_{11}=
\al^2(1-e^{-f_0}) + \be^2(1-e^{-f_0-f_1}),
\label{eq:S11}\\
\mathrm{Diffractive\,\, excit.}:&&T(|1,0\rangle \rightarrow |1,1\rangle) = 
  -S_{21}= \al \be \, e^{-f_0}(1-e^{-f_1}).
\label{eq:S21}
\end{eqnarray}

We note here in particular, that these results agree with the expressions in 
eqs.~(\ref{eq:eikonalel}, \ref{eq:eikonaldiff}), for the case
with $c_{ik}=U(-\delta, -\infty)$ given by eq.~(\ref{eq:toy1U}) and $T_i=1-(U_{\mathrm{int}})_{ii}$ (no
summation) given by eq.~(\ref{eq:toy1Uint}). 

\begin{figure}
  \begin{center}
    \includegraphics[scale=0.8]{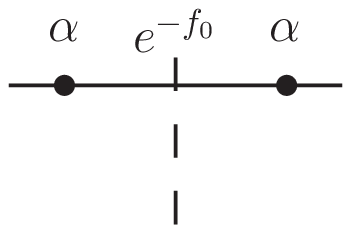}
    \hspace{1cm}
    \includegraphics[scale=0.8]{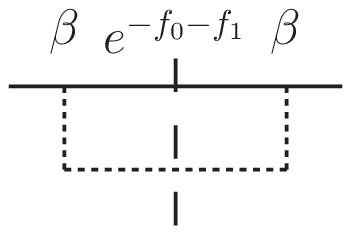}
  \end{center}
  \caption{The diagrams for elastic scattering in a two particle system. In
    this and subsequent figures a
    dashed line represents a virtual emission, which is reabsorbed by its
    parent. A dot represents a possible emission or absorption, which did not
    occur. The long dashed line indicates the interaction with the target,
    with the value for $U_{\mathrm{int}}$ indicated.} 
  \label{fig:toy1a}
\end{figure}

It will be helpful to represent the different contributions to 
the amplitudes by associated diagrams. Thus the elastic amplitude is 
interpreted as the sum of the two diagrams in
fig.~\ref{fig:toy1a}. In the first diagram no gluon is emitted before the
collision (weight $\al$), followed by the weight for no absorption of
the valence particle (weight $e^{-f_0}$), and finally no
emission is allowed after the interaction (weight $\al$). In the second
diagram a virtual gluon is emitted (weight $\be$), followed by the weight for
no absorption of the two-particle system (weight $e^{-f_0-f_1}$), and 
finally the virtual gluon has
to be reabsorbed by the valence particle (weight $\be$).

\begin{figure}
  \begin{center}
    \includegraphics[scale=0.8]{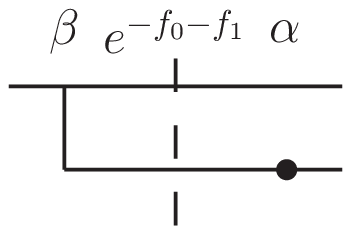}
    \hspace{1cm}
    \includegraphics[scale=0.8]{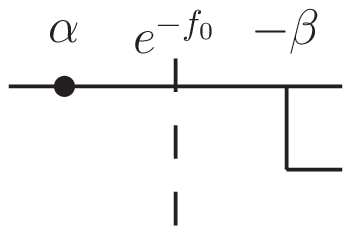}
  \end{center}
  \caption{The diagrams for diffractive excitation in a two particle system.}
  \label{fig:toy1b}
\end{figure}

In the same way the amplitude for diffractive excitation is represented by the
diagrams in fig.~\ref{fig:toy1b}. Here the emitted gluon appears as a real
particle in the final state. 
In the first diagram the gluon is 
emitted before interaction with weight $\beta$, then the system avoids
absorption by the target with weight $e^{-f_0-f_1}$, and finally the gluon
also avoids reabsorption by its parent with weight $\alpha$. In the second 
diagram the gluon avoids emission
before interaction with weight $\alpha$, the system is not absorbed by the
target with weight $e^{-f_0}$, and finally the gluon is emitted after the
interaction with weight ($-\beta$). This possibility is a consequence of the
unitarity of the evolution operator $U(-\delta,-\infty)$ in 
eq.~(\ref{eq:toy1U}), and the minus sign comes because it is now
emitted by $U^\dagger$ and not by $U$.
The two contributions interfere destructively, which gives the factor
$(1-e^{-f_1})$, which also can be interpreted as the amplitude for elastic
scattering of the emitted gluon.

This simple example illustrates two essential
features of the amplitudes for exclusive states in
diffractive excitation:

1. \emph{Fluctuations} in the scattering process are necessary for diffractive
excitation, as seen in 
eq.~(\ref{eq:eikonaldiff}). In this toy model the probability for gluon emission
is equal to $\beta^2$, so the average number of emitted gluons
hitting the target is $\langle n\rangle=\beta^2\cdot 1$, and the variance is 
$\langle n^2\rangle - \langle n\rangle^2 = \beta^2\cdot 1^2
-(\beta^2\cdot 1)^2=\alpha^2\beta^2$. Thus the factor $\alpha \beta$ in the 
amplitude in
eq.~(\ref{eq:S21}) (to be squared in the cross section) describes the 
fluctuations in the emission process. The last factor is the difference
between the interaction amplitudes of the states with and without emission,
and eq.~(\ref{eq:S21}) shows that diffractive excitation requires fluctuations
between different cascades with different interaction amplitudes. We also see
that 
diffractive excitation vanishes in the black disk limit, where $f_0$ and $f_1$
become very large, and diffractive scattering becomes purely elastic.

2. Diffraction is fundamentally a quantum mechanical process, where 
\emph{interference} is important. The two contributions in
fig.~\ref{fig:toy1a} or \ref{fig:toy1b} have to be
added in the amplitude, and not in the cross section.

It should also be noted that the time $t$ can be replaced by any variable, 
which  parametrizes the evolution between the incoming mass eigenstates and 
the eigenstates for the interaction. A BFKL based cascade would use the rapidity $y$, while a DGLAP cascade would prefer the transverse momentum $k_\perp$. We will here keep the notation $t$, and change to $y$ in sec.~\ref{sec:dipole}, where we will apply the model in \dipsy.

\subsubsection{Cascade with two possible emissions}
\label{sec:toy2b}

We now study a system in which two different gluons can be emitted. The gluons 
are denoted 1 and 2, and there are then
four different states, with one, two or three particles:
\begin{equation}
|1 0 0\rangle, \qquad |1 1 0\rangle, \qquad |1 0 1\rangle, \qquad |1 1 1\rangle. \label{eq:basis2}
\end{equation}
The two numbers are here occupation numbers for the three possible particle
states. 
  
We first study a sequential cascade, where the first gluon can be
emitted from the valence particle, and the second gluon can be emitted
from the first one. Thus the emissions are assumed to be ordered in
the evolution variable, and the second gluon can only be emitted if
the first one already is present. Note that we here do not consider
the possibility of a gluon is emitted from one parton and then reabsorbed
by another. Such recombination effects beyond the large $N_c$ limit
are discussed in secs.~\ref{sec:MCsat} and \ref{sec:swing}.

The evolution operator has now two components: the first, $U_1$, allows 
the first gluon to be emitted or absorbed by the valence particle, and the
second, $U_2$, allows the second gluon to be emitted or reabsorbed by gluon 1.
In the basis in eq.~(\ref{eq:basis2}), these operators can be written
\begin{equation}
U=U_2U_1,\quad\mathrm{where}\quad
U_1 = \left( \begin{array}{cccc}
\al_1 & -\be_1 & 0    & 0      \\
\be_1 & \al_1  & 0    & 0      \\
0     & 0     & \al_1 & -\be_1 \\
0     & 0     & \be_1 & \al_1  \\
\end{array} \right), \quad
U_2 = \left( \begin{array}{cccc}
1  & 0     & 0 & 0     \\
0  & \al_2 & 0 & -\be_2 \\
0  & 0     & 1 & 0     \\
0  & \be_2 & 0 & \al_2  \\
\end{array} \right). \nonumber
\end{equation}
Thus the first gluon can be emitted with probability $\beta_1^2$, and the
second with probability $\beta_2^2$, once the first is present. This asymmetry
also implies that the two matrices $U_1$ and $U_2$ do not commute.

The three particles have absorption probabilities $1-e^{-2f_0}$, $1-e^{-2f_1}$, and
$1-e^{-2f_2}$ respectively. The evolution operator describing the
interaction with the target is then given by
\begin{equation}
U_{\mathrm{int}} = \left( \begin{array}{cccc}
e^{-f_0} & 0 & 0    & 0      \\
0 & e^{-f_0-f_1}  & 0    & 0      \\
0     & 0     & e^{-f_0-f_2} & 0 \\
0     & 0     & 0 &e^{-f_0-f_1-f_2}   \\
\end{array} \right).
\label{eq:uint2}
\end{equation}
As in the earlier example, the evolution after the interaction is given
by the inverse of $U_2U_1$. Thus the projection on diffractive states of the $S$-matrix is
\begin{equation}
S=U_1^\dagger U_2^\dagger U_{\mathrm{int}} U_2 U_1.
\label{eq:Stoy2}
\end{equation}
Multiplying the matrices given above, and looking at the first column in 
the matrix $T=\mathbb{I}-S$, 
we get the amplitudes relevant for an incoming single valence particle:
\begin{eqnarray}
T_{\mathrm{el}}=T(|1 0 0\rangle \rightarrow |1 0 0\rangle)&=&\al_1^2(1-e^{-f_0}) 
+ \be_1^2 \al_2^2(1-e^{-f_0-f_1}) \nonumber\\ 
&&+\be_1^2\be_2^2(1-
e^{-f_0-f_1-f_2})\label{eq:Ttoy31}\\
T(|1 0 0\rangle \rightarrow |1 1 0\rangle)&=&e^{-f_0}\al_1\be_1 
[\al_2^2(1- e^{-f_1})+\be_2^2(1- e^{-f_1-f_2})]\label{eq:Ttoy32} \\
T(|1 0 0\rangle \rightarrow |1 0 1\rangle)&=&e^{-f_0}\be_1^2e^{-f_1}\al_2\be_2 (1
-e^{-f_2})\label{eq:Ttoy33} \\
T(|1 0 0\rangle \rightarrow |1 1 1\rangle)&=&e^{-f_0}\al_1\be_1e^{-f_1}\al_2\be_2 (1 -e^{-f_2}).
\label{eq:Ttoy34}
\end{eqnarray}

\begin{figure}
\begin{center}
  \includegraphics[scale=0.8]{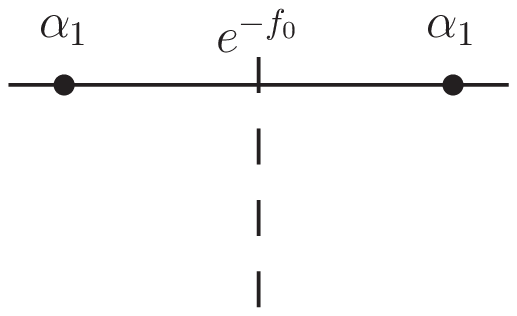}
  \hspace{0cm}
  \includegraphics[scale=0.8]{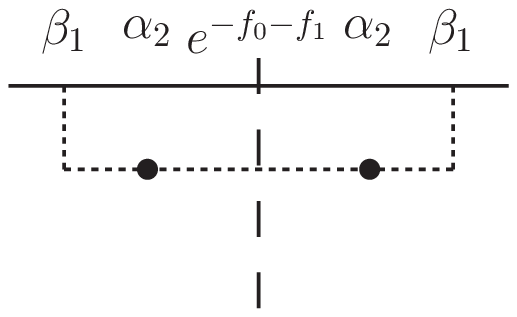}
  \hspace{0cm}
  \includegraphics[scale=0.8]{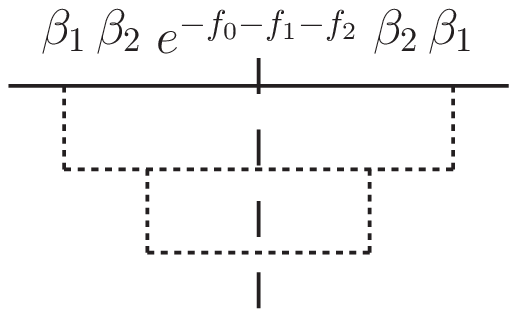}
\end{center}
\caption{\label{fig:toy2corr0} The diagrams for elastic scattering ($|100\rangle \rightarrow |100\rangle$) in an ordered three gluon system.}
\end{figure}

\begin{figure}
\begin{center}
  \includegraphics[scale=0.8]{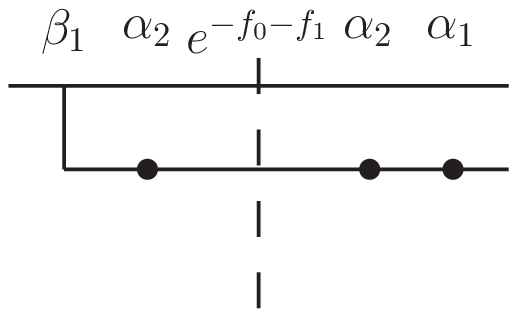}
  \hspace{0cm}
  \includegraphics[scale=0.8]{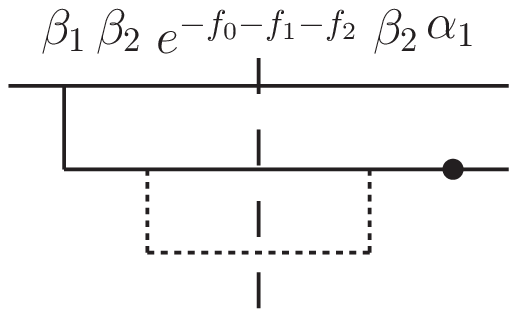}
  \hspace{0cm}
  \includegraphics[scale=0.8]{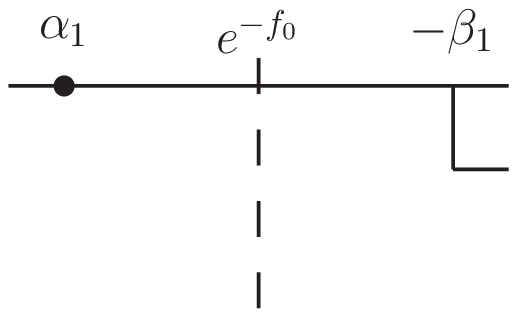}
\end{center}
\caption{\label{fig:toy2corr1} The diagrams for diffractive excitation of the first emission ($|100\rangle \rightarrow |110\rangle$) in an ordered three gluon system.}
\end{figure}

\begin{figure}
\begin{center}
  \includegraphics[scale=0.8]{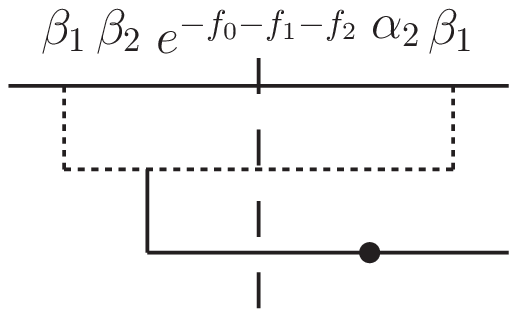}
  \hspace{1cm}
  \includegraphics[scale=0.8]{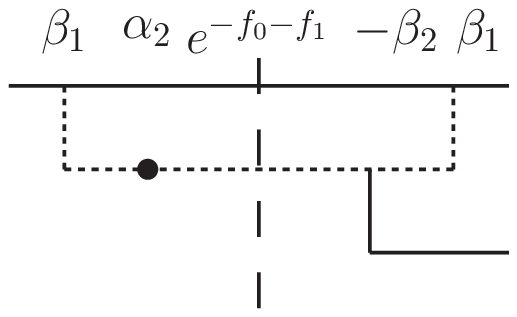}
\end{center}
\caption{\label{fig:toy2corr2} The diagrams for diffractive excitation of the second emission ($|100\rangle \rightarrow |101\rangle$) in an ordered three gluon system.}
\end{figure}

\begin{figure}
\begin{center}
  \includegraphics[scale=0.8]{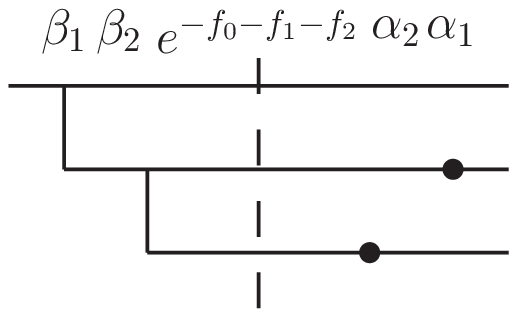}
  \hspace{1cm}
  \includegraphics[scale=0.8]{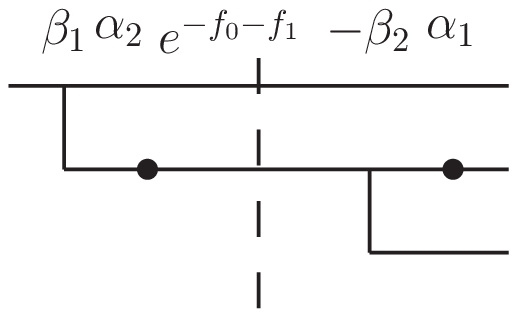}
\end{center}
\caption{\label{fig:toy2corr3} The diagrams for diffractive excitation of both emissions ($|100\rangle \rightarrow |111\rangle$) in an ordered three gluon system.}
\end{figure}

As in the previous example, these results can also be obtained from a sum 
of diagrams, as illustrated in figs.~\ref{fig:toy2corr0} -- \ref{fig:toy2corr3}. In these diagrams the
emission before, or the absorption after, the interaction gives a factor 
$\beta$,
while an emission after interaction gives  $-\beta$. The absence of a possible 
emission or reabsorption gives a factor $\alpha$. The interaction with the
target gives a factor $e^{-F}$, where $F=\sum f_i$, with the sum running
over the gluons present at the interaction. The sum of the contributions from
the relevant diagrams give the $S$-matrix elements corresponding to the
amplitudes in eqs.~(\ref{eq:Ttoy31} -- \ref{eq:Ttoy34}).


\begin{figure}
\begin{center}
  \includegraphics[scale=0.8]{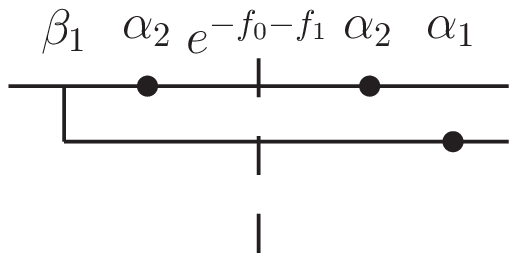}
  \hspace{1cm}
  \includegraphics[scale=0.8]{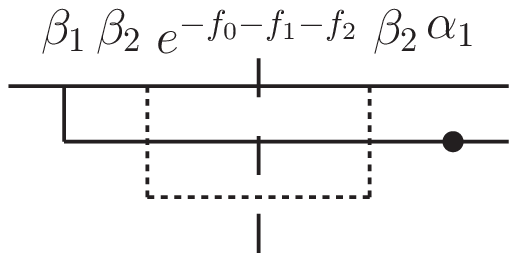}

  \includegraphics[scale=0.8]{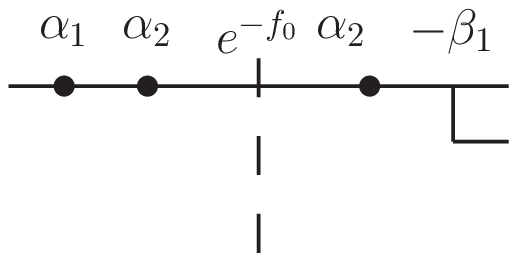}
  \hspace{1cm}
  \includegraphics[scale=0.8]{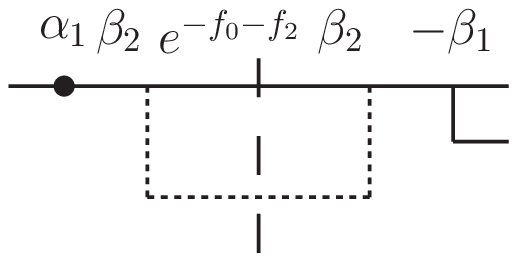}
\end{center}
\caption{\label{fig:toy2unc1} The diagrams for diffractive excitation of one out of two ($|100\rangle \rightarrow |110\rangle$) independent emissions.}
\end{figure}
\begin{figure}
\begin{center}
  \includegraphics[scale=0.8]{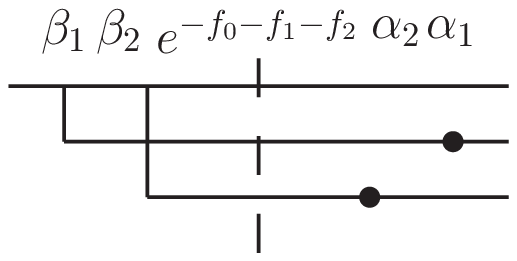}
  \hspace{1cm}
  \includegraphics[scale=0.8]{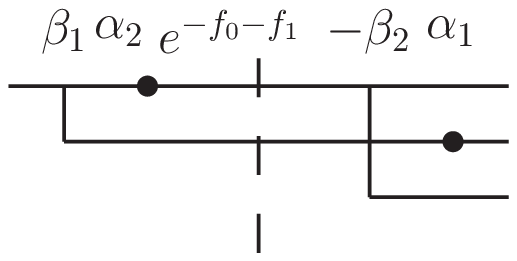}

  \includegraphics[scale=0.8]{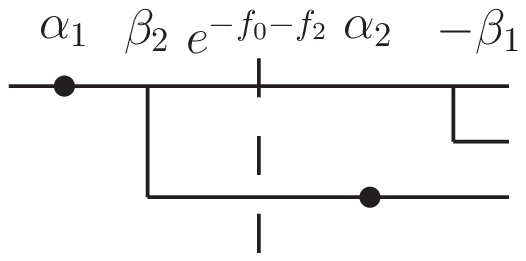}
  \hspace{1cm}
  \includegraphics[scale=0.8]{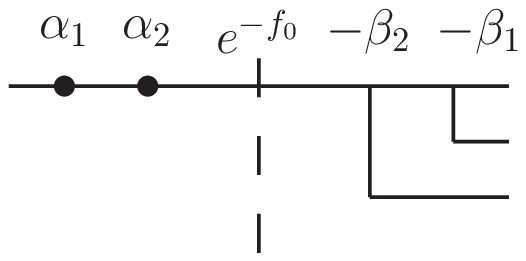}
\end{center}
\caption{\label{fig:toy2unc2} The diagrams for diffractive excitation of two ($|100\rangle \rightarrow |111\rangle$) independent emissions.}
\end{figure}

In case of two gluons, which can be \emph{independently} emitted from the valence
quark, the excited state $|110\rangle$, with a single emission, can be created through the four
diagrams in fig.~\ref{fig:toy2unc1}, which add up to
\begin{equation}
T(|1 0 0\rangle \rightarrow |1 1 0\rangle)
=e^{-f_0}\al_1\be_1(1-e^{-f_1})(\al_2^2+\be_2^2e^{-f_2})
\label{eq:Ttoy22}
\end{equation}
The transition to state $|111\rangle$, with both gluons emitted, would in this case also have four contributions, as shown
in fig.~\ref{fig:toy2unc2}. The result would then be
\begin{equation}
T(|0 0 0\rangle \rightarrow |111\rangle)=
e^{-f_0}\al_1\be_1\al_2\be_2(1-e^{-f_1})(1-e^{-f_2}).
\label{eq:Ttoy21}
\end{equation}
\subsubsection{Interpretation}

The results in eqs.~(\ref{eq:S11}, \ref{eq:S21}) and
(\ref{eq:Ttoy31}--\ref{eq:Ttoy21}) can be represented by
a few rules, which in sec.~\ref{sec:cont} and Appendix A are shown to hold also 
for a general cascade. 

\begin{enumerate}
\item 
A final state is specified by the \emph{real partons} 
present in the final state. It will also be important to separate 
those real emissions, which are the \emph{last gluon} in their chain,
from those, called \emph{parent gluons}, which have emitted at least one 
other real gluon.

The different diagrams, contributing to the
amplitude, may also contain ``virtual'' gluons, which are emitted but
reabsorbed in the cascade. Let $c_{V_i}^2$ denote the probability for emission
of a cascade $V_i$ from the real parton $i$. The virtual cascades can contain
more than one emission, and including no emission as representing an empty 
cascade, conservation of probability implies that $\sum_{V_i} c_{V_i}^2=1$ for
each $i$. The sum is over all possible virtual cascades emitted from the parton $i$.

\item
The contribution from a \emph{last} real parton $i$ is given by
\begin{equation}
\alpha_i \beta_i \sum_{V_i} c_{V_i}^2(1-e^{-f_i-F_{V_i}}).
\label{eq:LV}
\end{equation}

\item
The contribution from a \emph{parent} real parton $i$ is given by
\begin{equation}
\alpha_i \beta_i\sum_{V_i} c_{V_i}^2 e^{-f_i-F_V}.
\label{eq:PV}
\end{equation} 
\end{enumerate}

\emph{Comments}:
\vspace{2mm}

1) All \emph{real} emissions come with a factor $\alpha_i \beta_i$. Either
they are emitted before the interaction (weight $\beta_i$) and not reabsorbed
afterwards (weight $\alpha_i $), or not emitted before (weight $\alpha_i $)
and then emitted afterwards (weight $-\beta_i$).

2) If \emph{no virtual emissions} were possible, every \emph{last} gluon gives
a factor $(1-e^{-fi})$. This is the result
from interference between contributions from emission before and after the
interaction, and can also be interpreted as the amplitude for elastic
scattering from the target. 
Examples are gluon 1 in eq.~(\ref{eq:S21}) and
fig.~\ref{fig:toy1b}, and gluon 2 in eq.~(\ref{eq:Ttoy34}) and
fig.~\ref{fig:toy2corr3}.

3) When virtual emissions are possible, the factor $(1-e^{-fi})$ for a last
  gluon is modified. An example is the
amplitude $T(|1 0 0\rangle \rightarrow |1 1 0\rangle)$ in eq.~(\ref{eq:Ttoy32}),
where gluon 2 can be emitted and reabsorbed by gluon 1,
with probability $\beta_2^2$. Gluon 2 can only be emitted if gluon 1 is
emitted before the interaction, and therefore only affects the last term in the
factor $(1-e^{-f_i})$, changing it to $(1-e^{-f_1-f_2})$. Here the suppression
factor $e^{-f_2}$ represents the weight for \emph{no} absorption of gluon 2
from the target. (The cross section contains the factor $e^{-2f_2}$, which is
the probability for no absorption.) The probability for not emitting gluon 2 
(or for emitting an
empty cascade) is $\alpha_2^2=1-\beta_2^2$, and in this case there is no 
modification of the weight $(1-e^{-f_1})$. Adding the two contributions can be 
expressed as the sum in eq.~(\ref{eq:LV}).

4) The \emph{parent} gluons
cannot be emitted after interaction (they would then not be able to
emit their daughters), and therefore come with a
factor $e^{-f_i}$, which can be interpreted as the weight for 
not being absorbed by the target. The effect of a
virtual cascade $V$, is to reduce the weight for no absorption by a factor
$e^{-F_V}$, as expressed in eq.~(\ref{eq:PV}).

5) These results work also for the original valence particle, if it is treated
as a ``last'' particle for the elastic amplitude, and as a ``parent'' particle
in diffractive excitation, where it has emitted at least one real gluon. 
(There is no factor $\al_0\be_0$, as the valence particle is always present.) 
The difference is only that the two
terms in the parenthesis in eq.~(\ref{eq:LV}) do not correspond to emissions
before and after interaction. It is here a contribution from the definition
$T=\mathbb{I}-S$, which gives an extra term 1 in the elastic amplitude, reflecting 
the interference between the incoming and the scattered waves.
With this interpretation, an example of the sum over virtual emissions from a
parent particle is given by gluon 2 in 
eq.~(\ref{eq:Ttoy22}) and fig.~\ref{fig:toy2unc1}.

\subsection{Continuous cascade}
\label{sec:cont}

We now consider a cascade were every gluon can emit further gluons at every
point in the evolution parameter $t$. This corresponds to an infinite product
of evolution operators $U(t)$ with infinitesimal $\be_i(t)$.
In this subsection the gluons are assumed to interact individually, with 
no effects of screening from neighbouring
gluons. Such screening effects are taken into account in the dipole cascade 
discussed in sec.~\ref{sec:dipole}. We will here also assume that the cascade 
only includes gluon emission, and that effects 
of saturation from gluons joining within the cascades can be neglected.

The infinitesimal values for $\be_i$ implies that the corresponding values for
$\al_i$ are very close to 1, but as there is an infinite set of $\al_i$, the
small differences from 1 cannot be neglected. However, as the constraint
$\al_i^2+\be_i^2=1$ represents a conservation of probability, and therefore the
$\al$:s can be thought of as Sudakov factors, these corrections are automatically 
taken into account in a MC event generator like DIPSY. 

This probabilistic interpretation 
is also applicable for the emission of virtual cascades $V_i$, emitted from a
real gluon $i$. Let $v$ denote the individual gluons in the
virtual cascade $V$, and $\bar{v}$ the gluons which \emph{could} have been
emitted, either from gluon $i$ or from the gluons in $V_i$, but were not. In a
general situation the
probability for the gluon $i$ to emit the cascade $V_i$ is then given by
\begin{equation}
c_{V_i}^2= \prod_{v \in V_i} \be_{v}^2 \prod_{\bar{v} \in \bar{V}_i}
\al_{\bar{v}}^2. 
\label{eq:cv}
\end{equation}
Here $\be_{v}^2$ is the weight for emission and reabsorption of the gluon
$v$ in $V_i$, while $\al_{\bar{v}}^2$ is the weight for \emph{no}
emission of gluon $\bar{v}$ in $\bar{V}_i$ (where $\bar{V}_i$ is the set of all
possible emissions not included in $V_i$). 
As in the toy models discussed above, the sum of
emission probabilities for all possible cascades from the parent $i$,
adds up to one, 
\begin{equation}
\sum_{V_i} c_{V_i}^2 = \sum_{V_i} 
\left(\prod_{v \in V_i} \be_{v}^2 \prod_{\bar{v} \in \bar{V}_i}
  \al_{\bar{v}}^2\right)=1,
\label{eq:cvsum}
\end{equation}
as a result of the unitarity relationship between $\al$ and $\be$. In the sum
over all cascades $V$ we here also include the ``empty cascade'' $V_0$, with no 
emissions, for which  $c_{V_0}^2= \prod_{\bar{v} \in \bar{V}_i} \al_{\bar{v}}^2$.

In Appendix A we demonstrate that the rules presented in the previous section
also are relevant for a general cascade, and the 
amplitude for the transition to a real state $|R\rangle$ is given by
\begin{equation}
T_\All = \left(\prod_{\all \in \All} \be_\all \al_\all\right)
\left(\prod_{p\in P}\sum_{\casv} c_\casv^2 e^{-f_\cas-F_\casv}\right) 
\left(\prod_{\last \in \Last}  \sum_{\lastv}
  c_\lastv^2(1-e^{-f_\last-F_{\lastv}})\right).
\label{eq:full1}
\end{equation}
Here $\All$ is the set of all ``real gluons''. $V_p$ is a virtual cascade emitted 
from a gluon $p$ in the set of ``parent gluons'', denoted $\Cas$. $l$ is a gluon
in the set $L$ of ``last gluons'', and $V_l$ is a virtual cascade emitted from 
$l$. The sums over $V_p$ and $V_l$ run over all such cascades, including the 
empty cascades. Finally $F_{V_p}$ and $F_{V_l}$ denote the sum
of the relevant terms $f_i$. 

The probalastic interpretation of the factors $c_V^2$ in eqs.~(\ref{eq:cv},
\ref{eq:cvsum})
imply that this result can be generated by a MC event generator. The 
result can be interpreted as the probability to create the real state, times the
probability that  none of the parents is 
absorbed in the interaction with the target, times the probability that
the last gluon of each branch in the cascade should interact
elastically.

In this section the gluons are assumed to interact individually, with 
no effects of screening from neighbouring
gluons. Such screening effects are taken into account in the dipole cascade 
discussed in the following section. We have also assumed that the target does
not fluctuate like a parton cascade. Two colliding cascades will be discussed
in sec.~\ref{sec:target}.

\subsection{From gluons to dipoles}
\label{sec:dipole}

\subsubsection{Differences from the individual gluon cascade}

In the gluon cascade, discussed in the previous section, the interaction of a
gluon with the target is not changed by the emission of daughter gluons. The
gluons interact individually, with no effects of screening from neighbouring
gluons. Such screening effects are taken into account in the dipole cascade, 
where the screening is determined by the compensating charge in the other end
of a dipole. Thus a gluon in a small dipole has a smaller cross section,
leading to colour transparency. When a dipole is emitting a gluon, the
old dipole disappears, and is replaced by two new dipoles, as discussed in
section~\ref{sec:dipcasc}. This is also the case when a virtual 
cascade is emitted, which implies that
some of the real dipoles are no longer present in the interaction with the
target. Therefore the interaction amplitude $F$ cannot as easily be separated 
in one part from the real dipoles and one from the virtual emissions.
Some details in the calculations are rather technical and therefore left for
appendix A. Here we present first a toy model in
sec.~\ref{sec:dipoletoy},  which illustrates the
problem, and then a sketch of the full result in sec.~\ref{sec:dipolefull}.

\subsubsection{Toy model with single emission}
\label{sec:dipoletoy}
\begin{figure}
  \begin{center}
    \includegraphics[scale=0.8]{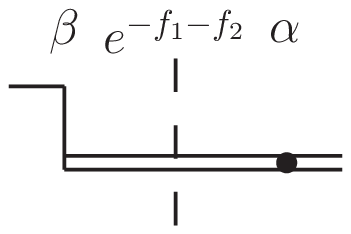}
    \hspace{2cm}
    \includegraphics[scale=0.8]{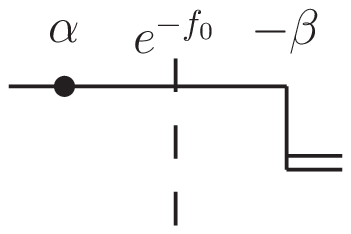}
  \end{center}
  \caption{\label{fig:diptoy} The amplitude for the transition $|100\rangle
    \rightarrow |011\rangle$ contains two contributions. The dipole can split
    by gluon emission before (left) or after (right) the interaction with the
    target.}
\end{figure}

We consider a simple toy model with two states, an initial single dipole
and a state where this is split in two daughter dipoles. Denoting the states
by the occupation numbers of the three dipoles, the single dipole state is
called $|1 0 0\rangle$ and the split state with two new dipoles is called $|0
1 1\rangle$. The weight for no absorption is $e^{-f_0}$, $e^{-f_1}$, and
$e^{-f_2}$ for the three dipoles. The diagrams contributing to diffractive
excitation are illustrated in fig.~\ref{fig:diptoy}.
To get a $|0 1 1\rangle$ final state, the emission can happen before or after
interaction, which gives two contributions to the amplitude. Thus we get
\begin{equation}
T(|100\rangle \rightarrow |011\rangle)=-S(|100\rangle \rightarrow |011\rangle) = -[\al \be e^{-f_1-f_2} + \al (-\be) e^{-f_0}] = \al \be (e^{-f_0}-e^{-f_1 - f_2})
\label{eq:toydipole}
\end{equation}
We note here that the weight $e^{-f_0}$ cannot be factored out in the same way 
as in eq.~(\ref{eq:S21}).
It will turn out that the last dipole to split, that is the ``second last
dipole'', will play a similar role also in the general case, discussed in the
next section.


\subsubsection{Full dipole cascade}
\label{sec:dipolefull}
\begin{figure}
  \begin{center}
    \includegraphics[width=0.5\linewidth]{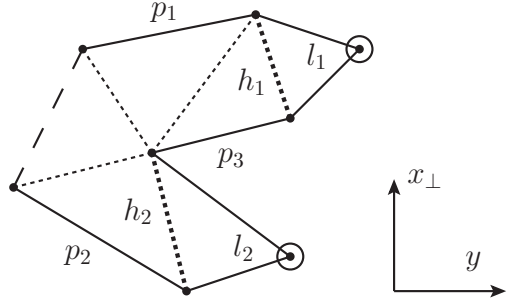}
  \end{center}
  \caption{\label{fig:dipcascade} A sample dipole cascade with rapidity on the
    horizontal axis and transverse space on the vertical axis. Full lines are
    the dipoles left in the final state, while dotted or dashed lines split
    into new dipoles before the final state. The circled gluons are the ones
    without children, and are the only gluons which can be emitted both before 
    and after interaction. The emission of a childless gluon creates a pair of
    dipoles denoted $l_i$. The childless gluons are emitted by the ``hidden'' 
    dipoles $h_i$. The dipoles  $\cas_i$ are the final state dipoles
    connected only to gluons with children. Dipoles that always split before
    interaction are marked 
    with thin dotted lines, and the long dashed line is the original dipole.}
\end{figure}

First the notation, illustrated in fig.~\ref{fig:dipcascade}, has to be 
slightly revised, as the interesting sets of
dipoles are different in some aspects. As in the previous section,
the set of \emph{real} dipoles in the final state is denoted $\All$. 
Each branch of a dipole cascade also has a \emph{last} 
gluon, which can be emitted before or after the interaction, while the rest 
of the emissions all have to occur before the interaction. In 
fig.~\ref{fig:dipcascade} the \emph{last gluons} in the cascade are circled.
Each last gluon is connected to two dipoles, formed when the last gluon was
emitted. The set of all such dipole \emph{pairs} is denoted $\Last$.
(Note also that two different last pairs can never have a common dipole.)
As in the cascade of independent gluons, the remaining real dipoles will be
called \emph{parent} dipoles
(included in the set $\Cas$), although this notation is not fully adequate. 
They are not parents in the sense that they have split into daughter dipoles.
They do, however, connect two gluons, which both have been involved in further
emissions. 

In the cascade with independent gluons in sec.~\ref{sec:cont}, the last emission is the addition 
of one more gluon, and the previous gluons remain unchanged. As discussed 
above, in a dipole
cascade the last emission will add two new dipoles, and \emph{remove} the
emitting dipole. The dipoles which
are split in the last emission play a special role, and are called
\emph{hidden} dipoles. In diagrams where the last split occurs \emph{before}
the interaction, the hidden dipoles do not contribute to the weight for no absorption from
the target, but they \emph{do} contribute if the last emission happens
\emph{afterwards}. This feature is illustrated in the toy model discussed above.

Virtual cascades in the dipole formalism will also remove the dipole it is
emitted from, while adding more dipoles. Therefore the emission of virtual
cascades from the ``parent dipoles'' do not give the factorizing non-interaction
weight $\exp(-F_p-F_{V_p})$ obtained in eq.~(\ref{eq:full1}). Instead we get 
when the cascade $V_p$ is non-empty, the weight $\exp(-F_{V_p})$, while for the
empty cascade we get $\exp(-F_p)$. We therefore here 
change the notation, and include the empty cascade in the set of all virtual
cascades, and define $F_{V_0}\equiv F_p$.
This definition implies that the non-interaction weight for the dipole $p$,
including possible virtual emissions, is always given by $\exp(-F_{V_p})$.

We use this notation also when a ``last dipole pair'' $l$ emits a virtual 
cascade $V_l$. Thus, when one or both of the dipoles in the pair $l$ is 
``emitting'' an empty cascade, the non-interaction weight $\exp(-F_{V_l})$
gets a contribution from the still intact dipole(s) in the pair~$l$.

Finally, in case the last dipole pair is produced \emph{after} the interaction,
virtual cascades can also be emitted by the hidden dipoles $h$. As these
cascades must be reabsorbed before the last dipole can be emitted, they are
restricted to the rapidity range $y>y_l$, where $y_l$ is the rapidity of the
last gluon. Also here the
non-interaction weight $\exp(-F_{V_h})$ is defined to equal $\exp(-F_{h})$,
when the cascade $V_h$ is the empty cascade.
With this notation the result in eq.~(\ref{eq:full1}), is for
a dipole cascade replaced by the following expression: 
\begin{equation}
T_\All = \left(\prod_{\all \in \All} \be_\all \al_\all \right)
   \left(\prod_{p\in P}\sum_{V_p} c_{V_p}^2 e^{-F_{V_p}}\right) 
   \left(\prod_{l\in L}\sum_{V_l} c_\lastv^2 \sum_{V_h (y>y_l)} c_{V_h}^2  
  (e^{-F_{V_h}}-e^{-F_{V_l}})\right). \label{eq:fulldip}
\end{equation}
Note here that there is no product over hidden dipoles, as the 
hidden dipole $h$ is specified by the last pair $l$.
The last factor, $(e^{-F_{V_h}}-e^{-F_{V_l}})$, is a generalization of the
result in the simple toy model.
The details of the calculations are presented in appendix A.

Note that, like for the independent gluon cascade in
eq.~(\ref{eq:full1}), in the derivation of these results it is assumed
that the virtual cascade from one emission is independent of the
virtual cascade from another emission. Thus the result corresponds to
a cascade without saturation effects from gluons which can join, or
dipoles which swing. These effects are further discussed in
secs.~\ref{sec:MC} and \ref{sec:swing}.

\subsection{Target cascade}
\label{sec:target}

\begin{figure}
  \begin{center}
    \includegraphics[scale=0.8]{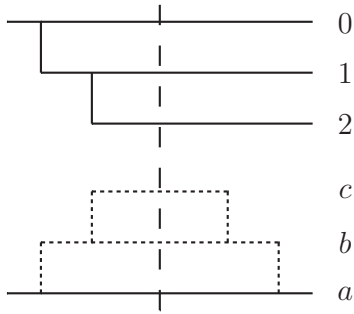}
  \end{center}
  \caption{\label{fig:targetcascade} A diagram contributing to single
    diffractive excitation in a collision between two cascades, where the
    target is scattered elastically. Labeling the partons in the upper
    projectile state 0, 1, and 2, and labeling the partons in the target cascade
    $a$, $b$, and $c$, with $a$ the incoming parton, the amplitude
    corresponding to this diagram is $\be_1\be_2 \be_b \be_c\exp
    (-F) \al_1 \al_2 \be_b \be_c$, with
    $F=f_{0a}+f_{1a}+f_{2a}+f_{0b}+f_{1b}+f_{2b}+f_{0c}+f_{1c}+f_{2c}$.}
\end{figure}

Up to now the target has been described as a potential without internal
structure. It can however also be described by a superposition of virtual
cascades, $V_t$,
in the same way as the projectile. For single diffractive excitation,
where the target is scattered elastically, the sum
over virtual cascades from the target particle should be summed over at
amplitude level (\emph{cf} eq.~(\ref{eq:eikonalel})). The amplitude can thus 
be written as an average over target configurations $V_t$, with weights
$c_{V_t}^2$, where all interaction amplitudes $F$ now depend on the target
cascade. For a dipole cascade we then get the amplitude
\begin{equation}
T_\All = \sum_{V_t} c_{V_t}^2 \left(\prod_{\all \in \All} \be_\all \al_\all\right) 
\left( \prod_{p\in P} \sum_{V_p} c_\Casv^2 e^{-F_{V_p,V_t}}\right)
\left(\prod_{\last \in \Last} \sum_{V_l} c_{V_l}^2 \sum_{V_h(y>y_l)} c_{V_h}^2 
(e^{-F_{V_h,V_t}}-e^{-F_{V_l,V_t}})\right). \label{eq:doubledip}
\end{equation}
$F_{X,T}$ is here the interaction amplitude between the virtual projectile
cascade $X$ and the target virtual cascade $T$, given as a sum
$F_{X,T}=\sum_{i,j} f_{i,j}$ over dipoles $i$ in $X$ and $j$ in $T$.

\section{Implementation in \dipsy}
\label{sec:MC}
The amplitude in eqs.~(\ref{eq:fulldip}, \ref{eq:doubledip}) can be used to generate diffractive events with the \dipsy event generator~\cite{Flensburg:2011kk}. The approach will be to first select a real cascade with the weight $\prod_{\all \in \All} \be_\all^2 \al_\all^2$, as it will appear in the squared amplitude, and then perform the sum over virtual cascades to determine the amplitude for this real state.

\paragraph{Real Cascade.}
First note that in a continuous cascade, each $\be$ is small, and each individual $\al$ is very close to 1. So the weight for the real cascades is essentially $\prod_{\all \in \All} \be_\all^2$. 
This weight does not include the probability that no further cascade is emitted after $\All$. To get the correct weight, any time the Monte Carlo generates a certain cascade $\All$ it should count towards the probability for that final state, no matter if the cascade goes on to generate more gluons afterwards. This is done by first generating a cascade $D$ with \dipsy as normal up to a maximum rapidity $y_{\text{max}}$, and return a random sub-cascade $\All \subset D$, weighted by the total number of sub-cascades of $D$.

As discussed in sec.~\ref{sec:cross-sections}, the structure of inelastic
final states is determined by the ``backbone'' or ``$k_\perp$-changing''
emissions. Presuming that diffraction is the shadow of absorption to inelastic 
states, we here conjecture that also diffractive states are specified by their
$k_\perp$-changing backbone gluons. Therefore real sub-cascades which contain
softer, not $k_\perp$-changing, emissions
have to go through the same reweighting procedure as the non-diffractive final
states, as discussed in sec.~\ref{sec:cross-sections} and fully described in
ref.~\cite{Flensburg:2011kk}. 

\paragraph{The sum over virtual cascades.}
For single diffractive excitation the sums over virtual cascades from the 
final excited state $R$ all come with
weights $c^2$, and can be calculated directly from \dipsy.
The sum over target cascades $\sum_{V_t}$ is independent of the excited state, and these cascades can be pre-generated and reused for each real state to save cpu-time.

For each real state, the subsets $\Cas$ and $\Last$ are identified, and
virtual cascades $V_p$, $V_l$, and $V_h$ are generated. Each of the 
pre-generated target cascades $V_t$ is paired up with a virtual cascade, and 
the factors $e^{-F_{V_p,V_t}}$ and $(e^{-F_{V_h,V_t}}-e^{- F_{V_l,V_t}})$ in the
amplitude in eq.~(\ref{eq:doubledip}) are calculated and multiplied together.
 The average over all the virtual cascades is then 
squared, and added to the weight for the real cascade 
($\prod_{\all} \be_\all^2$) discussed above, to give the total weight for the real state.


\paragraph{Saturation effects.}
\label{sec:MCsat}
In the Lund cascade model saturation effects are included within the cascade
evolution via the dipole ``swing'' (see sec.~\ref{sec:lundcascade}), in
addition to the effect from multiple subcollisions. 
The result in eq.~(\ref{eq:doubledip}) is valid for a linear cascade. In the
MC implementation, swings are included in each virtual cascade from each subset $P$, $h$, $l$, but swings between these virtual cascades are not considered in the current implementation. Swings are also included in the virtual cascade $V_t$ from the target.

In DIS, this omission is expected to be of negligible importance. First each
real chain end $\last$ gives a small factor $(1-e^{F_{V_l,V_t}})^2$, and
therefore events with two or more chain ends (corresponding to multiple
pomeron exchange) are suppressed, unless $Q^2$ is very small. Further, a large
contribution to the cross section comes from real states with 0 or 1 emission
from the $q\bar{q}$ pair, and in those cases there are no parent emissions
$\Cas$, and there can be no swing between different subsets of the real
cascade. For states with two or more gluons, there are corrections for the
swing, but the $1/N_c^2$ suppression makes it small as long as the emissions
are not too many. To further minimize the correction, the maximum rapidity 
for the generated projectile cascades, $y_{\text{max}}$, is
chosen as close to the real state as possible, to give the virtual cascades
from the real state little room to give saturation effects. This is
compensated by a longer evolution of the elastically scattered target proton, 
which includes the swing.

In the case of a diffractively excited proton, saturation is a larger
effect as the gluon density is higher and because the average dipole size is
larger. Its effect is, however, reduced because diffractive excitation is
dominated by peripheral collisions with relatively few interacting
dipoles. For central collisions with large interaction probability,
diffractive scattering is dominantly purely elastic. A method to calculate the
amplitude taking full account of saturation 
is introduced in section~\ref{sec:swing}, but this is not implemented in the
present MC. 

\paragraph{FSR and hadronisation}
As in the application of the DIPSY MC for generation of non-diffractive final
states~\cite{Flensburg:2011kk}, final state radiation is included 
in the same way as for the Linked Dipole Chain 
model~\cite{Gustafson:1986db,Gustafson:1987rq}
 in \ariadne~\cite{Lonnblad:1992tz}, and finally hadronisation using  
the Lund string model~\cite{Andersson:1983jt,Andersson:1983ia} implemented 
in \textsc{Pythia8}~\cite{Sjostrand:2007gs,Sjostrand:2006za}.
\vspace{3mm}

With these steps it is possible to generate full exclusive single diffractive
final states from the \dipsy event generator. We want to stress that 
\emph{no new
parameters} are introduced, which can be tuned to experimental data. All
parameters were previously determined by total and elastic cross section as functions of $\sqrt{s}$, and to some extent by exclusive non-diffractive data
and elastic reactions. Final state radiation in \ariadne and hadronization in 
\pythia have been tuned to exclusive LEP data.

\section{Early results}

As emphasized above, the cross sections for diffractive excitation are in our
formalism, 
via the optical theorem, fully determined by the projectile and target
cascades, and their absorption into non-diffractive
inelastic reactions.  For \dipsy, the inelastic absorption is tuned to
inclusive observables and $pp$ minimum bias multiplicity
in~\cite{Flensburg:2011kk}, and thus no additional parameters are introduced
in the extension to diffractive final states. It should also be noted that
tuning to experiments was done with respect to $pp$ data only, the only
exception being the total $\ga^*p$ cross section (or equivalently $F_2$) as
function of $Q^2$. 

In this section we show some early results with low statistics as proof of 
concept. We will continue with improvements in the present version of 
the MC implementation, which
will facilitate predictions for higher collision energies and
excitation masses.

\subsection{DIS}

\begin{figure}
  \begin{center}
    \includegraphics[angle=-90,width=0.6\linewidth]{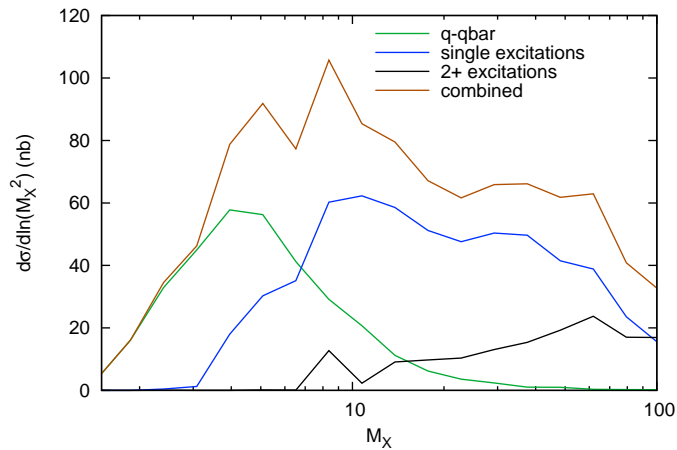}
  \end{center}
  \caption{\label{fig:dNdMX} The distribution in $\ln M_X^2$ for single
    diffractive excitation in DIS at $W=120$ GeV and $Q^2=24\,\mathrm{GeV}^2$.
    Besides the total result, also the contributions from $q\bar{q}$ states
    with no gluon emission, 1 gluon emission, and 2 or more gluon emissions
    are indicated. The distribution has a smooth cut off above 50 GeV due to the
    Lorentz frame used is the simulation.}
\end{figure}

Experimental results for diffractive excitation of the photon in DIS are 
available from HERA. Unfortunately a diffractively excited proton is mainly
outside the acceptance of the HERA detectors. Results for inclusive 
diffractive cross sections and for
$d\sigma_{SD} /d M_X^2$ could be easily obtained from the original DIPSY MC,
and were presented in ref.~\cite{Avsar:2007xg}, in good 
agreement with data from HERA. In the new formalism it is
also possible to obtain information about the partonic content of the
diffractive states.
 In fig.~\ref{fig:dNdMX} we show our result
for the distribution in $\ln M_X^2$ at $W=120$ GeV and
$Q^2=24\,\mathrm{GeV}^2$. The figure also shows the separate contributions
from states with 0, 1 and $\ge2$ gluons besides the $q\bar{q}$ pair initially
coupled to the virtual photon. We see that there is a bump at 
$M_X^2 \approx Q^2$, which corresponds to final states formed by the initial 
$q\bar{q}$ state. This bump is followed 
 by a more flat distribution, obtained from states including also one or more
 gluons. As expected, states with more gluons become increasingly important
 for higher masses $M_X$.
The distribution in the figure has a smooth cut off around 50 GeV, due to the
    Lorentz frame used in the simulation, which limits the rapidity range for
    the gluons in the excited state. 

\begin{figure}
  \begin{center}
    \includegraphics[angle=-90,width=0.49\linewidth]{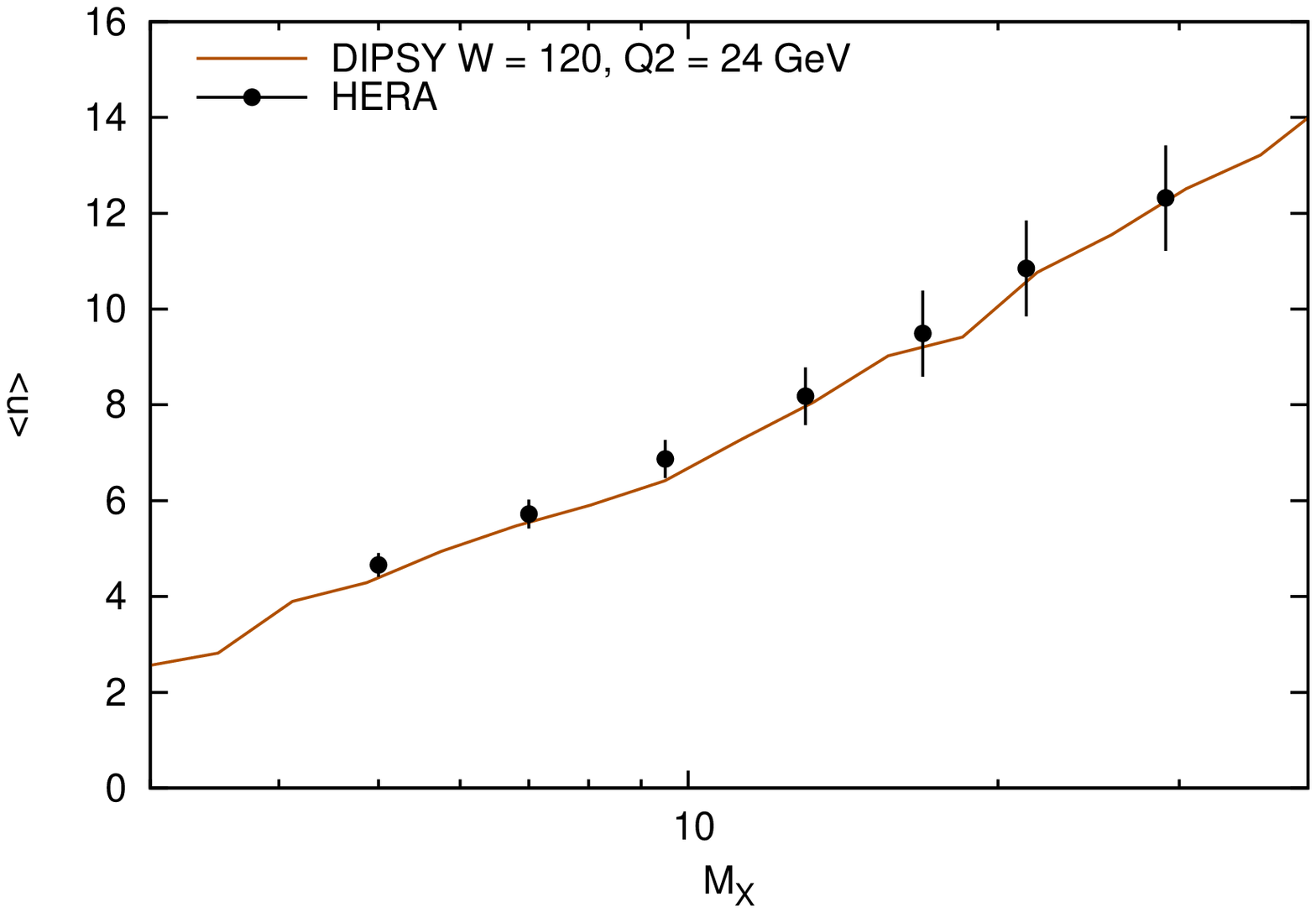}
    \includegraphics[angle=-90,width=0.49\linewidth]{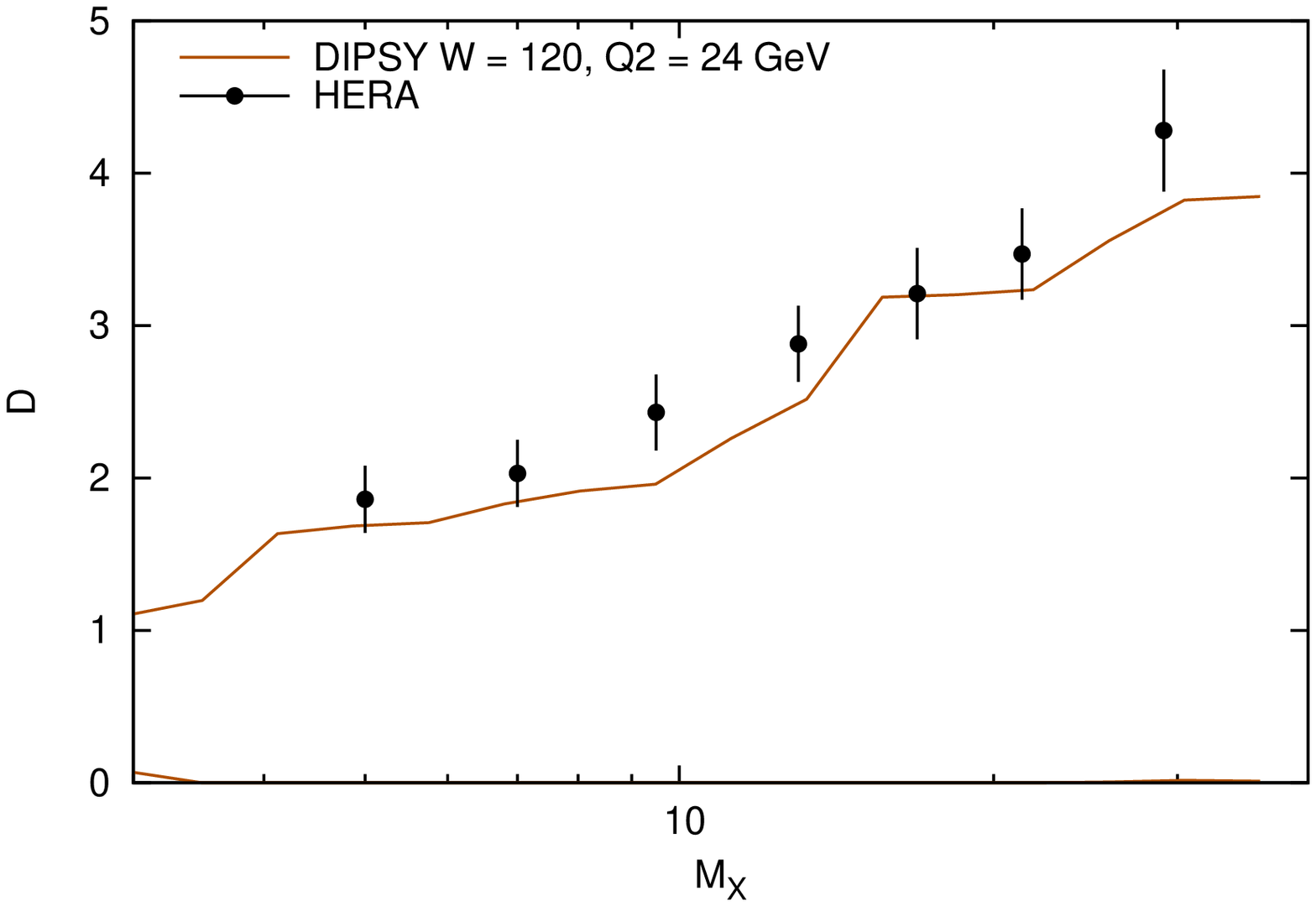}
  \end{center}
  \caption{\label{fig:avN} The average total charged multiplicity (left) and its fluctuations (right) as function of excited photon mass $M_X$ in single diffractive DIS at HERA~\cite{Adloff:1998dw}. }
\end{figure}

\begin{figure}
  \begin{center}
    \includegraphics[angle=-90,width=0.48\linewidth]{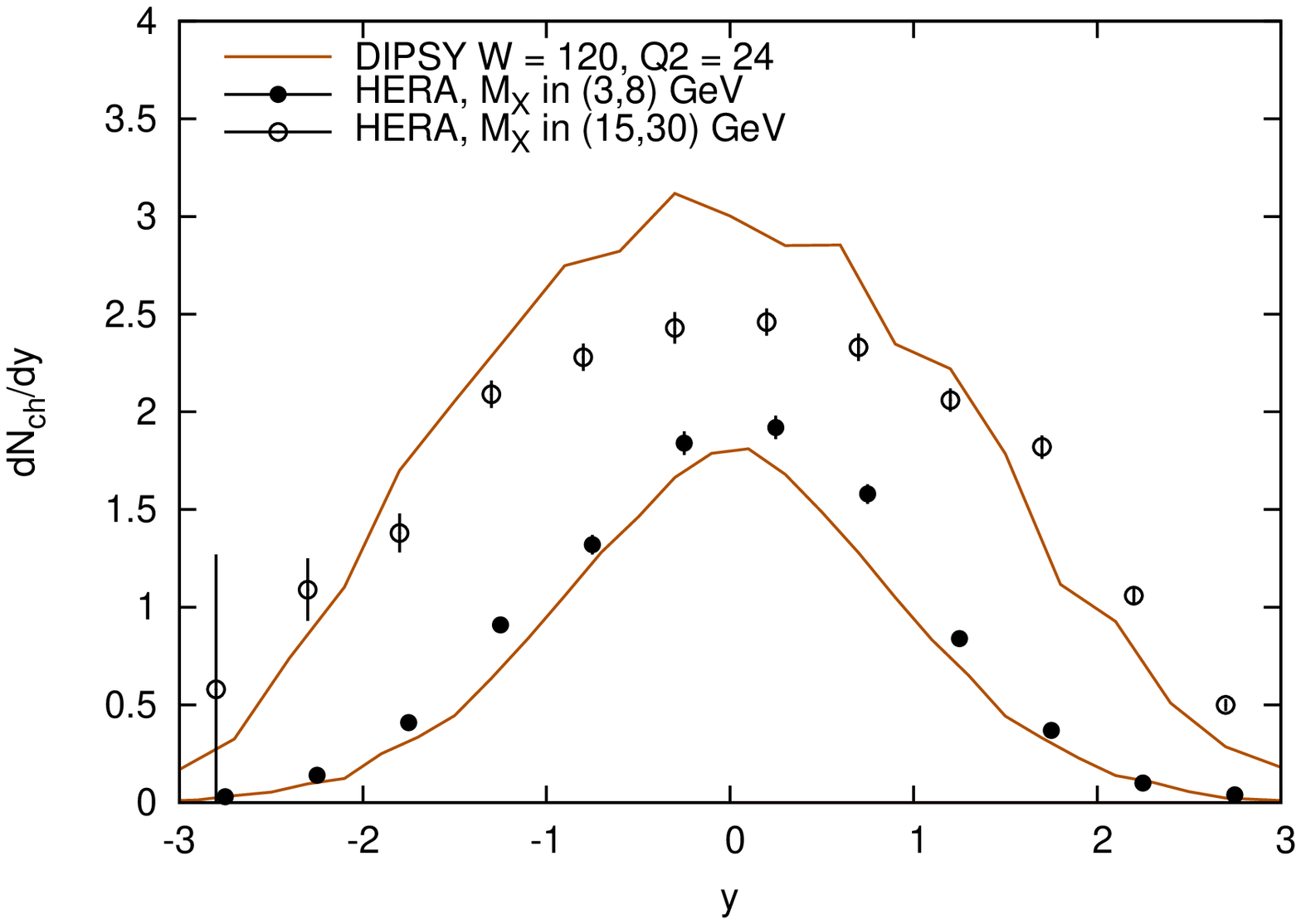}
    \includegraphics[angle=-90,width=0.48\linewidth]{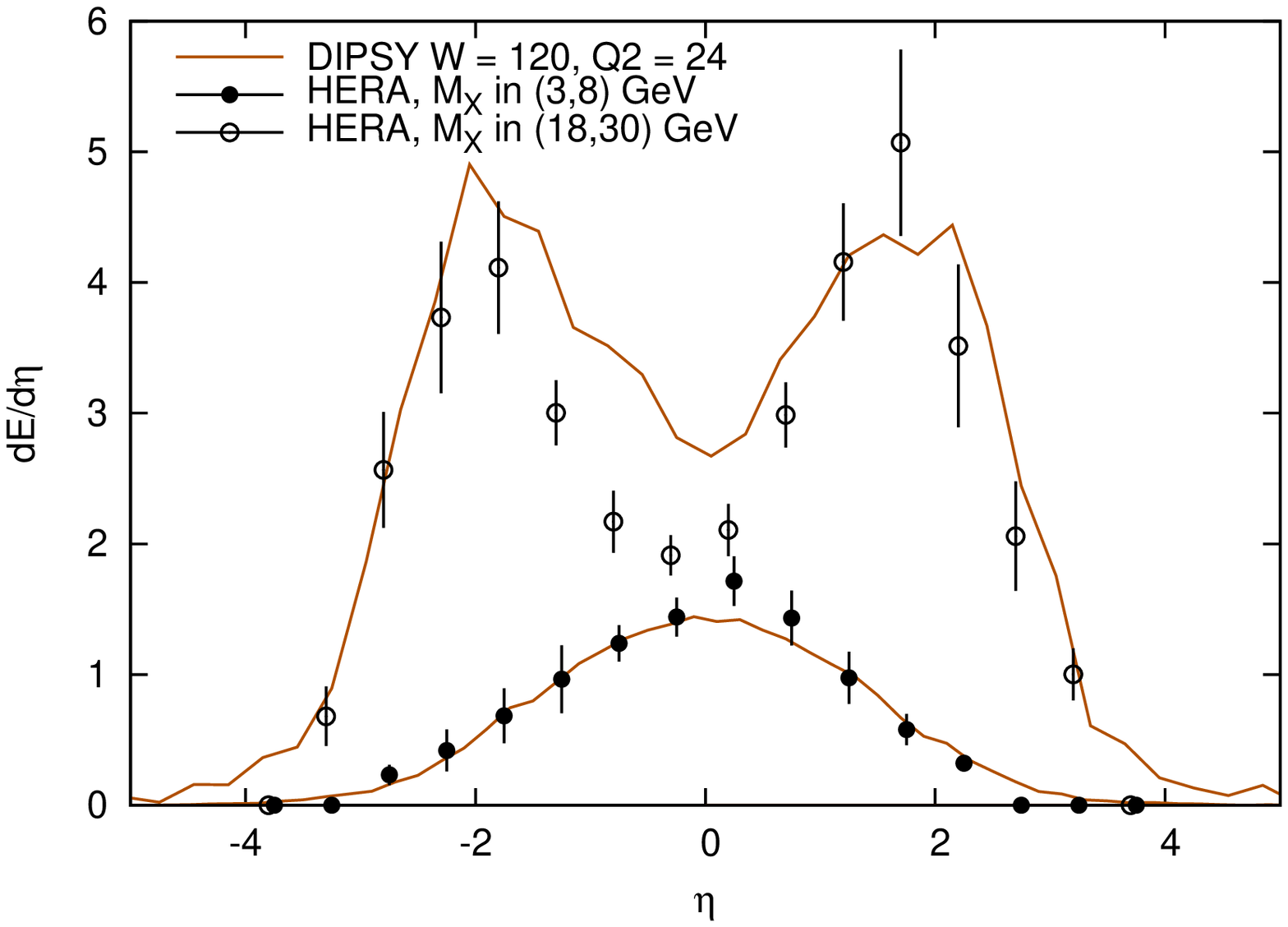}
  \end{center}
  \caption{\label{fig:etadep} The average charged multiplicity as function of 
    rapidity
    in bins of $M_X$ (left) \cite{Adloff:1998dw}, and average energy flow as
    function of pseudorapidity in bins of $M_X$ (right)
    \cite{Adloff:1998ds}. The results correspond to $W=120$ GeV and
    $Q^2=24\,\mathrm{GeV}^2$.} 
\end{figure}

Some results for the properties of the hadronic final state, obtained
after final state radiation and hadronization, are presented in
figs.~\ref{fig:avN} and \ref{fig:etadep}, and compared with HERA data
from H1~\cite{Adloff:1998dw, Adloff:1998ds}. In all cases the results
are obtained for $W=120$ GeV and $Q^2=24\,\mathrm{GeV}^2$. In
fig.~\ref{fig:avN} we show the average total charged multiplicity and
its fluctuations, as function of excited photon mass $M_X$ in single
diffractive excitation. We see here a very good agreement between our
results and the experimental data, both for the average and for the
fluctuations. Fig.~\ref{fig:etadep} shows the average charged
multiplicity as function of rapidity, and average energy flow as
function of pseudorapidity, in two bins of $M_X$. Also here we find a
rather good agreement with data. The fact that it agrees for both the
multiplicity and the energy flow, indicates that also the average
transverse momentum has to be correctly reproduced.

\subsection{$pp$ and $p\bar{p}$ collisions}

\begin{figure}
  \begin{center}
    \includegraphics[width=0.48\linewidth]{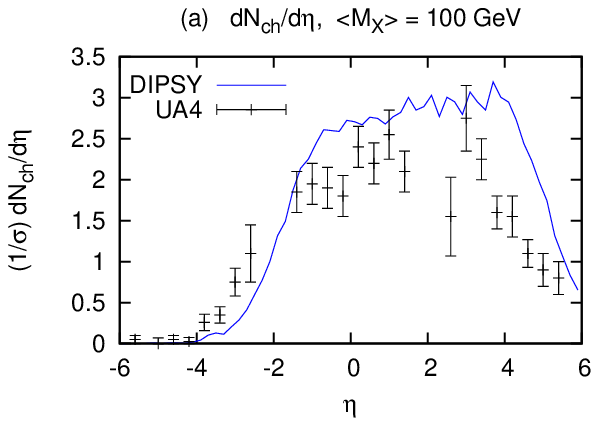}
    \includegraphics[width=0.48\linewidth]{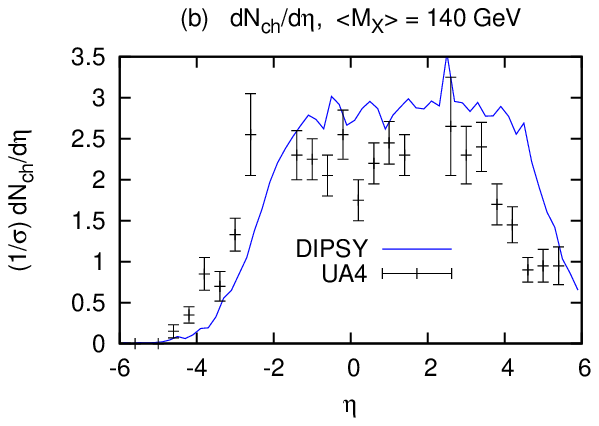}
  \end{center}
  \caption{\label{fig:UA4} Pseudorapidity density $dN_{ch}/d\eta$ (in the
    total cms) in single diffraction in two mass bins with 
    $\langle M_X\rangle=100$ GeV and
    $\langle M_X\rangle=140$, in $p\bar{p}$ collisions at $W=546$ GeV. The
    results from DIPSY are compared with data
    from UA4~\cite{Bernard:1985kh}. The excess of particles at large $\eta$ is
  related to the lack of a leading baryon in our model, see the main text.}
\end{figure}

Also in colliders for $pp$ or $p\bar{p}$ collisions most detectors 
have limited acceptance
for a diffractively excited proton. At the CERN $p\bar{p}$ collider, results
for distributions in pseudorapidity 
were presented from the UA5 and UA4 detectors \cite{Bernard:1985kh,
Ansorge:1986xq}. Fig.~\ref{fig:UA4} shows $\eta$-distributions
for single diffraction in two mass bins with $\langle M_X\rangle=100$ GeV and
$\langle M_X\rangle=140$, in $p\bar{p}$ collisions at $W=546$ GeV. We see that
a rapidity plateau with $d N_{ch}/d\eta \sim 2.5$ is developed for larger
$M_X$-values, in fair agreement with the experiment. We also note that the
simulation gives a surplus of particles at high $\eta$, amounting to
approximately one extra particle. This fact is related to the lack of a
leading baryon in our model. This baryon is expected to take a significant
fraction of the forward energy, and thus reduce the particle density for large
$\eta$-values. The lacking baryon is a consequence of our simple proton
wavefunction, which contains only gluons. This was motivated by the fact that
after a long cascade evolution, the resulting parton distribution at small $x$
is rather insensitive to the initial wavefunction. The particles at high
$\eta$ in fig.~\ref{fig:UA4} do, however, depend sensitively on the large $x$ 
partons.

\begin{figure}
  \begin{center}
  \epsfig{file=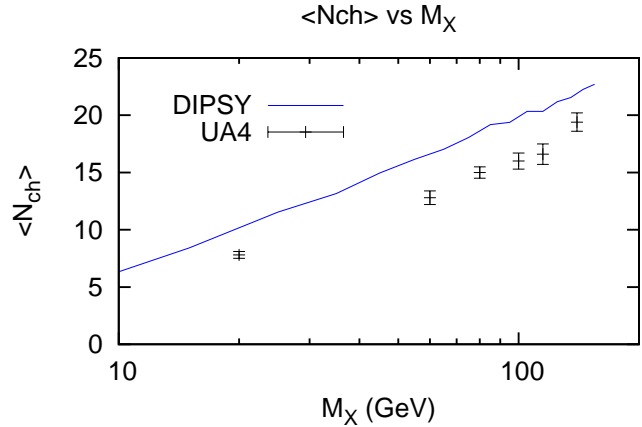,width=0.6\textwidth}
  \end{center}
  \caption{The average charged multiplicity as a function of $M_X$
    generated by DIPSY and compared to data from UA4
    \cite{Bernard:1985kh}.\label{fig:UA4AvNch}}
\end{figure}

The lack of quarks can also explain our general
overestimate of the charged multiplicity, which is almost constant
over a wide range of $M_X$, as can be seen in
fig.~\ref{fig:UA4AvNch}. Our diffractive
final states will always contain a double string connecting a leading
gluon in the backward direction with the remnant, while in reality
there will also be diffractive states with a leading quark, giving
rise to only one string and, therefore, approximately only half the
multiplicity. The fraction of such states should decrease with
increasing $M_X$ and the rise of the multiplicity is mainly determined
by gluonic components, but it may very well explain the constant
surplus multiplicity in our model. In addition, the fact that a
leading quark has a harder fragmentation function may partly explain
why our simulations undershoots data at large negative rapidities in
fig.~\ref{fig:UA4}.

The lack of quarks has also effects in other places, where large $x$-values
are important, \emph{e.g.} for jets with very high $p_\perp$,
 and in future improvements we will have to include quarks in the proton
wavefunction. 

\section{Future developments}
\label{sec:future}

We here discuss some effects and reactions, which are not implemented in
the present MC, but which can be included in future versions.

\subsection{Double diffraction}
\label{sec:double}

The results for single diffractive excitation, discussed in the previous
sections, can also be generalized to double diffraction. The result is,
however, numerically more complicated and the implementation in a MC would be
more time consuming.
For simplicity we here discuss the result for independent gluon
cascades, but the result can be directly generalized to dipole cascades.
Consider a diffractive reaction in which the projectile is excited to a
real cascade $R_{\mathrm{p}}$, and the target to a real cascade
$R_{\mathrm{t}}$. The number of ``last'' gluons is $n_{\mathrm{p}}$ and
$n_{\mathrm{t}}$ respectively. The last gluons can be emitted either
before or after the interaction, which implies that there are
$2^{n_{\mathrm{p}}+n_{\mathrm{t}}}$ different diagrams contributing to the
amplitude. For a specific diagram
the sets of last gluons emitted \emph{before} the interaction in the
projectile and target cascades are denoted $B_{\mathrm{p}}$ and $B_{\mathrm{t}}$.
Each pair of two last gluons in these sets, $b_{\mathrm{p}}$ from the
projectile and $b_{\mathrm{t}}$ from the target, contributes a weight
$\sum_{V_{b{\mathrm{p}}}}c^2_{V_{b{\mathrm{p}}}} \sum_{V_{b{\mathrm{t}}}}
c^2_{V_{b{\mathrm{t}}}} \exp(-F_{V_{b{\mathrm{p}}}, V_{b{\mathrm{t}}}})$.
Here the sums run over all virtual cascades emitted from $b_{\mathrm{p}}$
and $b_{\mathrm{t}}$. (We here use the notation introduced in 
sec.~\ref{sec:dipolefull}, and let $F_{V_{b{\mathrm{p}}}, V_{b{\mathrm{t}}}}$ include
contributions from the parent gluons $b_{\mathrm{p}}$ and $b_{\mathrm{t}}$.) 
In addition these gluons also interact with all
``parent'' gluons coming from the other side. The ``last'' gluons, which are
emitted after the interaction, do not take part in the interaction, and give only the factor
$(-1)^N$, where $N$ is the number of such gluons in the diagram.

The parent gluons also interact with each other in the same way as for
single diffraction. In obvious notation the total contribution from this
diagram is
\begin{eqnarray}
&&(-1)^N\,\left(\prod_{r_{\mathrm{p}}\in R_{\mathrm{p}}}
\beta_{r_{\mathrm{p}}}\alpha_{r_{\mathrm{p}}}\right)
\left( \prod_{r_{\mathrm{t}}\in
R_{\mathrm{t}}}\beta_{r_{\mathrm{t}}}\alpha_{r_{\mathrm{t}}}\right) \nonumber\\
&&\times\prod_{p_{\mathrm{p}}\in P_{\mathrm{p}}}
\sum_{V_{p_{\mathrm{p}}}}c^2_{V_{p_{\mathrm{p}}}}
 \prod_{p_{\mathrm{t}}\in P_{\mathrm{t}}}
\sum_{V_{p_{\mathrm{t}}}}c^2_{V_{p_{\mathrm{t}}}}
 \prod_{b_{\mathrm{p}}\in
B_{\mathrm{p}}}\sum_{{V_{b_{\mathrm{p}}}}}c^2_{V_{b_{\mathrm{p}}}}
 \prod_{b_{\mathrm{t}}\in B_{\mathrm{t}}}\sum_{V_{b_{\mathrm{t}}}}
c^2_{V_{b_{\mathrm{t}}}} \nonumber\\
&&\times\, e^{(-F_{V_{p_{\mathrm{p}}},V_{p_{\mathrm{t}}}}-
    F_{V_{p_{\mathrm{p}}},V_{b_{\mathrm{t}}}}-
    F_{V_{b_{\mathrm{p}}},V_{p_{\mathrm{t}}}}-
    F_{V_{b_{\mathrm{p}}},V_{b_{\mathrm{t}}}})}
\end{eqnarray}
Summing over the $2^{n_{\mathrm{p}}+n_{\mathrm{t}}}$ different diagrams
then gives the total amplitude for the double diffraction final state $R$.
The result does not simplify to a factorized expression as the result for single
diffraction in eq.~(\ref{eq:full1}), but it can 
still be calculated in a MC, only more time consuming.

\subsection{Full account of saturation effects}
\label{sec:swing}

In sec.~\ref{sec:finalstates} the cascade evolution was assumed to be
linear, that is, each gluon will radiate (or each dipole split) independently
from the rest of the cascade. This is a good approximation for $\ga^*p$, with
an excited photon with moderate or high $Q^2$, but in the case of $pp$
scattering, non-linear effects play a very important role in
diffractive scattering~\cite{Flensburg:2010kq}. Diffractive excitation is very
much suppressed in central collisions, where diffraction is dominated by
elastic scattering. Thus diffractive excitation is largest in peripheral
collisions, where saturation is not equally important.
In the present MC, saturation effects are, besides from multiple subcollisions
in the Lorentz frame used, also included
with swings within individual sub-cascades, as described in 
sec.~\ref{sec:MCsat}. We expect this to account for most of the saturation
effects in events with not too high masses $M_X$. However, for excitation to
higher masses swings between dipoles in \emph{different} cascades may also be
important, and we here  
discuss how it is possible to include these interactions, and thus take full
account of saturation effects. 

This problem is similar to the problem with double diffraction discussed above,
in that the sum of contributions from all possible diagrams does not factorize. 
Consider a real cascade with $n\geq 2$ branches ending in childless gluons
$l_i$. As described in sec.~\ref{sec:finalstates}, these gluons can be emitted
either before or after the interaction. This implies that there are $2^n$
different diagrams which contribute to the amplitude. Virtual sub-cascades can
be emitted from a childless gluon if this is emitted before interaction, but
not if it is emitted after the interaction. If the sub-cascades do not interact
with each other, the result factorizes as shown in eq.~(\ref{eq:full1}) or  
(\ref{eq:fulldip}). This is no longer the case if the evolution of one virtual
sub-cascade depends on whether another childless gluon is emitted before or
after the interaction with the target. This implies that all the $2^n$
different diagrams must be calculated separately, and added to give the
full production amplitude. This is not a problem in principle, but it becomes
numerically difficult as the virtual cascades have to
be calculated for each of the $2^n$ possible combinations of childless gluons
emitted before the interaction. It becomes very time consuming, unless an
efficient weighting system is introduced, which favours the generation of
states with a relatively high production probability. 

\subsection{Quarks in the proton wavefunction}

The model for a proton wavefunction in the MC contains only
gluons. The motivation is that gluons dominate the cascades at high
energy, and for evolutions down to small $x$, the result is rather
insensitive to the exact starting configuration. For diffractive
events this motivation is only relevant for excitation to large
masses. For lower $M_X$, experimental results are consistent with
particle production from a single string, stretched between a pulled
out quark and the proton remnant \cite{Smith:1985fa}. To describe
these states it is necessary to include quarks in the initial proton
wavefunction.  This problem will be addressed in future work.

\subsection{Formalism combining diffractive and non-diffractive events}
\label{sec:combined}
We have in this paper discussed diffractive events with one or more pomerons
exchanged. The pomerons are ladders formed by gluon pairs, which mediate
momentum exchange and thus are able to excite the projectile to higher
mass. In our earlier treatment of non-diffractive reactions
\cite{Flensburg:2011kk}, 
we have studied events where one or more dipole branches interact with the
target via gluon exchange. This exchange causes a colour connection between 
the projectile cascade and the target, and provides momentum which puts the
interacting dipole branch on shell. Branches which do not interact in this way
are treated as virtual, and reabsorbed in the projectile state. 

\begin{figure}
  \begin{center}
    \includegraphics[scale=0.8]{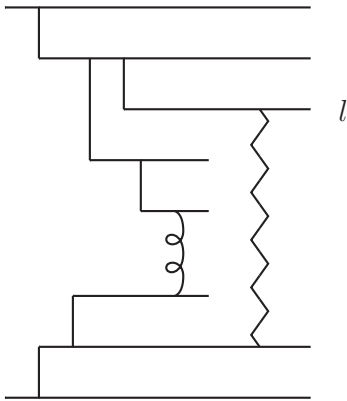}
  \end{center}
  \caption{\label{fig:combined} A cascade with both colourful and colour
    neutral exchange with the target. The spiral line represents exchange of 
    a single gluon with
    the target, which gives a colour connection. The zigzag line represents
    exchange of a colour neutral pomeron. It transfers momentum implying
    that the branch ending in the gluon marked $l$, can come on
    shell, but it does not result in a colour connection between this gluon and
    the target.} 
\end{figure}

We here discuss how it is possible to include exchange of both 
(uncut) pomerons and gluons in
a unified formalism. An example of an event with both types of exchange in a
single event, is shown in fig.~\ref{fig:combined}. There is no gap
in the event, but the
branch connected to the pomeron is coming on shell, increasing the
multiplicity. In the formalism in ref.~\cite{Flensburg:2011kk} the probability
for this final state is included in the non-diffractive cross section, but the
branch interacting with the pomeron is treated as virtual, and neglected in
the final state. Although not modifying the inclusive non-diffractive cross
section, this implies a small underestimate of the multiplicity in
these events. It would, however, not affect any results presented in this
paper. A further discussion about this combined formalism, and how it could be 
implemented in the event generator, is presented in appendix B.

\subsection{Hard diffraction and multiple gap events}

The present version of the DIPSY MC generates unbiased events. This implies
that it is quite inefficient for producing hard diffraction, like events with
high $p_\perp$ jets, for which some kind of theoretical trigger would be
needed. It also does not include multiple gap events. Here we believe that the
formalism described in sec.~\ref{sec:combined}, can be generalized to describe
events with several uncut pomerons, including diffractive events with more
than one gap, or double diffraction with overlapping diffractive systems.

\section{Conclusions}

In this paper we describe a formalism, in which exclusive final states in
diffractive excitation can be determined from basic QCD dynamics, BFKL
evolution and saturation.
Although diffraction is basically a quantum-mechanical phenomenon with strong 
interference effects, we show here how it is still possible to calculate the
different 
interfering components to the amplitude in a MC simulation, add them with
proper signs, and thus
calculate the reaction cross section for exclusive diffractive final states.

 Our formalism is based on the Good--Walker 
formalism for diffractive excitation, and 
it is assumed that the virtual parton cascades represent
the diffractive eigenstates defined by a definite absorption amplitude,
in analogy with refs.~\cite{Miettinen:1978jb, Hatta:2006hs, Avsar:2007xg,
  Flensburg:2010kq}.
Thus we here assume that events with large rapidity gaps can be
understood as analogous to diffraction in optics, with matrix elements
determined by absorption into inelastic channels via the optical
theorem. This is also the case for
the Regge formalism, where the calculation of diffractive cross
sections is based on the AGK cutting rules and Mueller's triple-Regge
formalism. In ref. \cite{Gustafson:2012hg} we also conjecture that the
Good--Walker and triple-pomeron formalisms actually are different
views of the same dynamical phenomenon. 

Also a description where gaps are the result of ``soft colour
reconnection'' in initially colour connected inelastic events, has
successfully described many experimental observations
\cite{Edin:1995gi, Pasechnik:2010cm}.  This is also the case for a
description of gap events in terms of projectile-pomeron collisions
\cite{Ingelman:1984ns}, which is implemented in a number of event
generators.  The true nature of rapidity gap events is therefore still
not fully revealed.  An essential feature of our scheme is here that
the result is fully determined by the inelastic reactions, with no
extra tunable parameters, pomeron flux factors or pomeron parton
distributions, which have to be tuned to data.

The formalism is based on the Lund dipole cascade model,
which in turn is based on BFKL evolution and saturation, and is
implemented in the DIPSY event generator \cite{Avsar:2005iz, Flensburg:2011kk}.
The model is a generalization of Mueller's dipole cascade model
\cite{Mueller:1993rr,Mueller:1994jq,Mueller:1994gb}, including also
non-leading effects and saturation effects within the evolution. 
It has previously been successfully applied to inclusive diffractive 
cross sections in DIS and $pp$ collisions~\cite{Avsar:2007xg,
  Flensburg:2010kq}, and later also to exclusive final 
states in non-diffractive events~\cite{Flensburg:2011kk}.
The assumption that the interaction is dominated by
absorption into inelastic channels implies that all contributions to the 
amplitudes are real, and therefore can be added with their relative signs, and
afterwards squared to give the relevant reaction cross sections.
This feature also implies that it is possible to calculate the Fourier
transform of the $b$-distribution, and determine the $t$-dependence (some
results are presented in ref.~\cite{Flensburg:2010kq}). 

We here note that, as discussed in
refs.~\cite{Andersson:1995ju,Salam:1999ft}, the structure of non-diffractive
final states is determined by the ``backbone'' or ``$k_\perp$-changing''
emissions. For exclusive final states, final state radiation has to be added
with Sudakov form factors, thus not changing the inclusive cross sections. 
Assuming that diffraction is the shadow of absorption to inelastic 
states, we here implicitly assume that also diffractive states are specified 
by their $k_\perp$-changing backbone gluons, where final state radiation and
hadronization must be added to get the detailed final states. 


We show in this paper some early results for single diffraction in DIS and $pp$
collisions, with comparisons to experimental data. 
As emphasized above, these results were obtained without introducing any new 
parameters to the model, and the predictions for exclusive diffractive 
observables are entirely determined by tuning to non-diffractive and elastic 
data. 

In the near future we plan to present more detailed analyses of the model,
include quarks in the proton wavefunction, 
and make predictions for LHC energies. We also want to compare our
results with those from other approaches, \emph{e.g.} 
\textsc{Pythia}~\cite{Sjostrand:2007gs} and SHRIMPS~\cite{Martin:2012nm}. 
In the somewhat longer perspective we may include double
diffraction, multiple gap events, and a formalism combining diffractive
and non-diffractive events, as described in sec.~\ref{sec:future}. This
development will need a significant improvement of the algorithms. Hard
diffraction would in addition need some kind of theoretical trigger.

\section*{Acknowlegments}

Work supported in part by the Swedish research council (contracts
621-2009-4076 and 621-2010-3326).

\section*{Appendices}
\appendix
\section{Final states for a full continuous cascade}
\label{sec:A}
We here derive the results for the transition amplitudes in
eqs.~(\ref{eq:full1},\ref{eq:fulldip}).  For simplicity we only discuss the
independent gluon cascade. The
modifications in a dipole cascade were discussed in sec.~\ref{sec:dipolefull}.

\begin{figure}
  \begin{center}
    \includegraphics[scale=0.8]{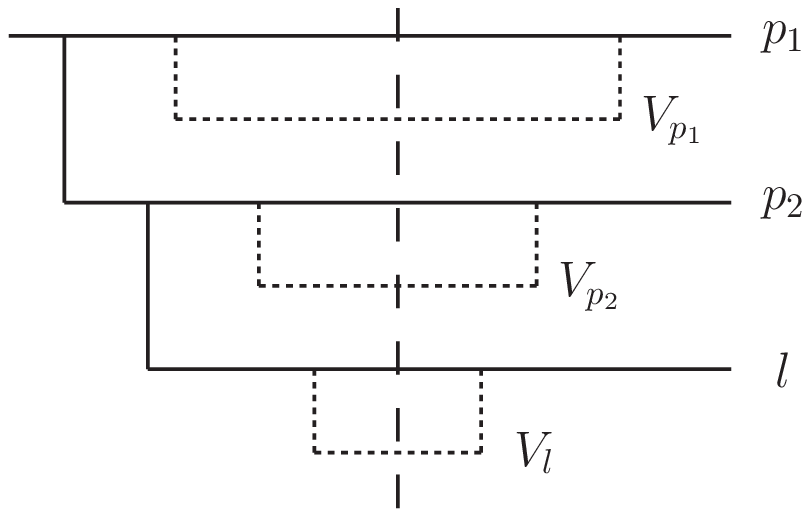}
    \hspace{0cm}
    \includegraphics[scale=0.8]{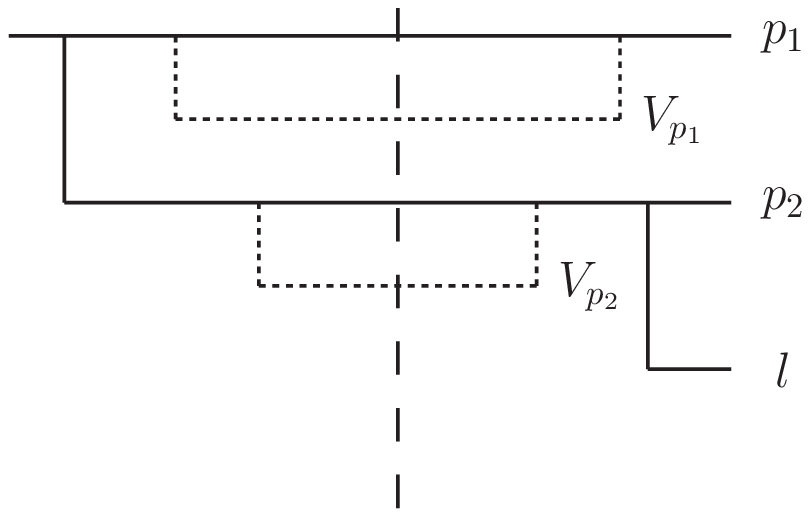}
  \end{center}
  \caption{\label{fig:contsing} The final state $R = P \cup l$ gets
    contributions  from diagrams where $l$ is emitted before interaction
    (left) and after interaction (right). Due to ordering, virtual cascades
    from $l$ (such as $V_l$) are only possible when $l$ is emitted before
    interaction.} 
\end{figure}

We first study the simpler case with a single real chain, where the set of
``last'' gluons, $\Last$,
contains only a single gluon $l$. This is the case illustrated in 
fig.~\ref{fig:contsing}. The emission structure of the real gluons $\All$
can then only be in two configurations:
either they are all emitted before interaction (left diagram of fig.~\ref{fig:contsing}), 
or the last gluon, $l$, is emitted afterwards
(right diagram of fig.~\ref{fig:contsing}). As discussed above, the ``parent''
gluons must always be emitted
before interaction, to give the last gluon a chance to be emitted. 

We begin with the diagram on the left of fig.~\ref{fig:contsing}, where the last real
gluon is emitted \emph{before the interaction}. This part of the $S$-matrix must
include a sum over all possible virtual cascades $V_i$, emitted
from the partons, $i$, in the real cascade. The emission probability for a particular virtual 
cascade $V_i$ is given by $c_{V_i}^2$ defined in eq.~(\ref{eq:cv}), which
satisfy the constraint $\sum c_{V_i}^2=1$ where the sum runs over all virtual
cascades for a fixed real parton $i$.
The contribution to the $S$-matrix from all diagrams of this type is given by
\begin{eqnarray}
S_B &=& \prod_{\all \in \All}  \left( \be_\all \al_\all 
\sum_{V_r} c_{V_r}^2  e^{-(f_r+F_{V_r})}\right)\nonumber\\
&=&\left(\prod_{\all \in \All} \be_\all \al_\all \right)
\left(\prod_{p\in P} \sum_{V_p} c_{V_p}^2 e^{-f_p-F_{V_p}} \right) 
\left( \sum_{V_l} c_{V_l}^2 e^{-f_l-F_{V_l}} \right).
\label{eq:SB}
\end{eqnarray}
Here $f_r$ and $F_{V_r}=\sum f_v$ denote the Born-level interactions, with
the sum running over all gluons in the cascade $V_r$. Thus the factor
$e^{-(f_r+F_{V_r})}$ represents the weight for no inelastic interaction by the
gluon $r$, or by its associated virtual cascade $V_r$. The sums in
eq.~(\ref{eq:SB}) run over all
possible virtual cascades, with their respective probabilities.
In the last line we have separated the product over real gluons into a product 
over parent gluons times a contribution from the last gluon $l$. We have also
pulled out the product over $\be_\all \al_\all$.

The second contribution to the $S$-matrix, where $l$ is emitted 
\emph{after the interaction} (see right diagram of fig.~\ref{fig:contsing}), 
contains an extra $(-1)$, and allows only virtual cascades from the ``parent''
gluons $p$. Thus this contribution is
\begin{equation}
S_A = (-1) \left( \prod_{\all \in \All} \be_\all \al_\all\right)
\left(\prod_{p\in P} \sum_{V_p} c_{V_p}^2 e^{-f_p-F_{V_p}} \right).
\label{eq:SA}
\end{equation}

Adding the contributions in eqs.~(\ref{eq:SB}, \ref{eq:SA}) gives the 
$S$-matrix element for transition from an incoming state $|0\rangle$
to an excited state $|R\rangle$. For a non-elastic transition
 we have $T=-S$, which gives the result
\begin{eqnarray}
T(|0\rangle \rightarrow |\All\rangle)&=&-(S_B + S_A) = 
\left(\prod_{\all \in \All} \be_\all \al_\all \right)
\left(\prod_{p\in P} \sum_{V_p} c_{V_p}^2 e^{-f_p-F_{V_p}} \right)
\left(1 - \sum_{V_l} c_{V_l}^2 e^{-f_l-F_{V_l}} \right) \nonumber\\
&=&\left(\prod_{\all \in \All} \be_\all \al_\all \right)
\left(\prod_{p\in P} \sum_{V_p} c_{V_p}^2 e^{-f_p-F_{V_p}} \right)
\left(\sum_{V_l}  c_{V_l}^2(1 - e^{-f_l-F_{V_l}}) \right).
\end{eqnarray}
In the last equality we have here used the relation $\sum c_{V_l}^2=1$.

In a general real cascade there may be several branches, which end in a 
``last'' gluon. 
As before, the parent gluons in the real cascade, which are not the
last ones in their branch, have to be emitted before the
interaction. Otherwise their children could not be emitted. As in the simpler
example above, they must all contribute a factor $(\sum_{V_p} c_{V_p}^2
e^{-f_p-F_{V_p}})$. The last gluons can all be emitted either before or after 
the interaction, and they will therefore all give a factor 
$(-\sum_{V_l}  c_{V_l}^2(1 - e^{-f_l-F_{V_l}}))$ to the $S$-matrix element. As
the sign of the transition amplitude is not important for the reaction rate,
we get for a cascade of independent gluons the final result (\emph{cf}
eq.~(\ref{eq:full1})) 
\begin{eqnarray}
|T(|0\rangle \rightarrow |\All\rangle)| = 
\left(\prod_{\all \in \All} \be_\all \al_\all \right)
\left(\prod_{p\in P} \sum_{V_p} c_{V_p}^2 e^{-f_p-F_{V_p}} \right)
\left(\prod_{l\in L}\sum_{V_l}  c_{V_l}^2(1 - e^{-f_l-F_{V_l}}) \right).
\label{eq:Scont}
\end{eqnarray}

\section{Events with both diffractive and non-diffractive subcollisions}

\label{sec:mixed}
In the Good-Walker formalism only diffractive states are considered. Cascades
which get colour-connected to the target are treated as absorbed, and removed
from the incoming wave. The coherent sum of cascades forming the proton
wavefunction is distorted when different components are absorbed with
different probabilities. This distortion, which leads to diffractive
excitation, is described by a non-unitary $S$-matrix, which is insensitive to
the nature of the absorbed states. 


Within the Lund dipole cascade formalism it is also possible to describe the
exclusive non-diffractive states responsible for the absorption
\cite{Flensburg:2011kk}, 
and it is possible to combine the two descriptions in a unified formalism,
which can be implemented in the event generator DIPSY. In such a scheme the 
interaction will be described by a unitary $S$-matrix.



\paragraph{Toy model}
To illustrate the method we first study the simple toy model discussed in 
sec.~\ref{sec:toy1}. In this model there is an initial valence particle, which 
is able to emit a gluon. There are
two different diffractive states, $|1,0\rangle$ and $|1,1\rangle$, where 
the numbers 0 and 1 denote an empty or
occupied parton state. We now extend the set of states, and let the number 2
indicate a parton which has interacted via gluon exchange, and is attached to
the target. Thus we have 6 different states:
\begin{equation}
 |1,0\rangle,\quad |1,1\rangle,\quad  |1,2\rangle,\quad |2,0\rangle,\quad
 |2,1\rangle,\quad |2,2\rangle.
\end{equation}

We have also to generalize the evolution operator $U(-\delta, -\infty)$ in 
eq.~(\ref{eq:toy1U}), which describes the emission of a gluon. The valence 
particle can emit a gluon with weight 
$\beta$, if the gluon state is empty. This is possible also if the valence
parton is attached to the target,  and we therefore have
\begin{equation}
U(-\delta, -\infty) |1,0\rangle =\alpha|1,0\rangle + \beta |1,1\rangle,\quad
 U(-\delta, -\infty) |2,0\rangle =\alpha|2,0\rangle + \beta |2,1\rangle.
\end{equation}
All other states already contain an emitted gluon, and are unaffected by 
$U(-\delta, -\infty)$.

The interaction with the target described by $U_{\mathrm{int}}$ has now,
besides the elastic non-interaction weights $e^{-f_0}$ and $e^{-f_1}$ in
eq.~(\ref{eq:toy1Uint}), also non-diffractive interactions transforming a free
valence parton to an attached one, with weight $\sqrt{1-e^{-2f_0}}$, and
similarly transforming a free gluon to an attached gluon with weight 
$\sqrt{1-e^{-2f_1}}$. This implies that $U_{\mathrm{int}}$ can be written in a
factorized form $U(-\de,+\de) =
U_{\mathrm{int}}=U_{\mathrm{int},0}U_{\mathrm{int},1}$, where 
\begin{equation}
U_{\mathrm{int},0} |1,i\rangle = \nn_0 |1,i\rangle + \yn_0 |2,i\rangle, \qquad 
U_{\mathrm{int},1} |j,1\rangle = \nn_1 |j,1\rangle + \yn_1 |j,2\rangle. 
\end{equation}
Here $\nn_i = e^{-f_i}$ is the non-absorption amplitude, and $\yn_i = \sqrt{1-e^{-2f_i}}$ the non-diffractive interaction amplitude. $U_{\mathrm{int},0}$ does not act on gluon 1 and vice versa. Note that
\begin{equation}
\nn_i^2 + \yn_i^2 = 1, \nonumber
\end{equation}
as now $U_{\mathrm{int}}$ is a unitary operator.

Finally the evolution after the interaction is given by $U(+\infty, +\delta)=
U^\dagger(-\delta, -\infty)$, which implies that the non-diagonal elements
have changed sign. (We have defined the phases such that $\alpha$
and  $\beta$ are real.) Note that since $U(-\delta, -\infty)$ can only emit
free gluons, an attached gluon cannot be absorbed by $U(+\infty, +\delta)$.
The unitary $S$-matrix is given by
$S=U^\dagger(-\delta, -\infty) U_{\mathrm{int}} U(-\delta, -\infty)$, and it
is straightforward to read off the transition amplitudes for an incoming
state with a single valence parton:

\begin{eqnarray}
S(|1,0\rangle \rightarrow |1,0\rangle) &=&  \nn_0 ( \al^2 + \be^2 \nn_1 ) = e^{-f_0} \left( \al^2 + \be^2 e^{-f_1} \right) \nonumber\\
S(|1,0\rangle \rightarrow |1,1\rangle) &=&  - \al \be \nn_0 (1 - \nn_1) =  -
\al \be e^{-f_0} \left( 1 - e^{-f_1} \right) \nonumber\\
S(|1,0\rangle \rightarrow |1,2\rangle) &=& \be \nn_0 \yn_1 = \be e^{-f_0} \sqrt{1-e^{-2f_1}} \nonumber\\
S(|1,0\rangle \rightarrow |2,0\rangle) &=& \yn_0 ( \al^2 +\be^2 \nn_1  ) = \sqrt{1-e^{-2f_0}} \left( \al^2 +\be^2 e^{-f_1}  \right)  \nonumber\\
S(|1,0\rangle \rightarrow |2,1\rangle) &=& - \al \be \yn_0 (1 - \nn_1) = - \al \be \sqrt{1-e^{-2f_0}} \unit{-f_1} \nonumber\\
S(|1,0\rangle \rightarrow |2,2\rangle) &=& \be \yn_0 \yn_1 = \be \sqrt{1-e^{-2f_0}} \sqrt{1-e^{-2f_1}} \nonumber
\end{eqnarray}

We note here that the amplitudes for elastic scattering and diffractive excitation,
in the first two lines, agree with the result in sec.~\ref{sec:toy1}. The
inclusive non-diffractive cross section also agrees with the eikonal result
$d \sigma_{\mathrm{inel}}/d^2 b =\langle 1-e^{-2F}\rangle=\alpha^2(1-e^{-2f_0})+
\beta^2(1-e^{-2f_0-2f_1})$, used in earlier publications.
However, in our analysis of exclusive non-diffractive final states in
ref.~\cite{Flensburg:2011kk}, the possibility for momentum exchange which can
put a virtual branch on shell, as the emitted gluon in state $|2,1\rangle$,
was not included. Thus in the present version of the DIPSY event generator 
the cross section for the state $|2,0\rangle$, with only the valence parton
connected to the target, is overestimated and given a probability representing
both states $|2,1\rangle$ and $|2,0\rangle$.

\paragraph{General cascade}
The above result generalises easily to a full cascade of multiple chains with
continuous emissions. For clarity we here show the result for cascades of
independent gluons, as discussed in sec.~\ref{sec:cont}. We let $N$ denote the 
non-diffractive cascade, consisting of the interacting gluons (set $I$) and
their ancestors (set $A$). As before $R$ is the set of gluons in the branches
which come on shell via pomeron exchange, with $L$ denoting the last gluons in
its branch and $P$ denoting their parents (including grandparents etc., but
not including any gluon in $N$). The generalization of eq.~(\ref{eq:full1}) 
then reads
\begin{eqnarray}
S(|0\rangle \rightarrow |N &+& \All\rangle) = 
\left\{\left( \prod_{n \in N} \be_n \right)
\left(\prod_{i\in I} \sqrt{1-e^{-2f_i}} \sum_{V_i} c_{V_i}^2 e^{-F_{V_i}}\right) 
\left(\prod_{a\in A}  \sum_{V_a} c_{V_a}^2 e^{-F_a-F_{V_a}}\right)
\right\}
\nonumber\\
&&\times\left\{ \left(\prod_{\all \in \All} \be_\all \al_\all\right)
\left( \prod_{p\in P}\sum_{V_p} c_{V_p}^2 e^{-F_p-F_{V_p}} \right)
 \left(\prod_{l\in L} \sum_{V_l} c_{V_l}^2 (1-e^{-F_l-F_{V_l}})
 \right)\right\}. 
\label{eq:mixedfull}
\end{eqnarray}
This result can be implemented in a future version of DIPSY, and would imply
slightly higher multiplicities in non-diffractive events.

\bibliographystyle{utcaps}
\bibliography{references,refs}

\end{document}